\documentclass[aps,prl,reprint,superscriptaddress,longbibliography]{revtex4-2}
\usepackage{blindtext}
\usepackage{graphicx}
\usepackage{amsfonts}
\usepackage{amsmath}
\usepackage{bbold}
\usepackage{physics}
\usepackage{float} 
\usepackage{xcolor}
\usepackage[export]{adjustbox}

\begin{document}
\title{Full Counting Statistics of Charge in Chaotic Many-body Quantum Systems}
\author{Ewan McCulloch}
\affiliation{Department of Physics, University of Massachusetts, Amherst, MA 01003, USA}
\affiliation{Department of Electrical and Computer Engineering,
Princeton University, Princeton, NJ 08544, USA}

\author{Jacopo De Nardis}
\affiliation{Laboratoire de Physique Th\'eorique et Mod\'elisation, CNRS UMR 8089,
CY Cergy Paris Universit\'e, 95302 Cergy-Pontoise Cedex, France}

\author{Sarang Gopalakrishnan}
\affiliation{Department of Electrical and Computer Engineering,
Princeton University, Princeton, NJ 08544, USA}

\author{Romain Vasseur}
\affiliation{Department of Physics, University of Massachusetts, Amherst, MA 01003, USA}

\begin{abstract}

We investigate the full counting statistics of charge transport in $U(1)$-symmetric random unitary circuits. We consider an initial mixed state prepared with a chemical potential imbalance between the left and right halves of the system, and study the fluctuations of the charge transferred across the central bond in typical circuits. Using an effective replica statistical mechanics model and a mapping onto an emergent classical stochastic process valid at large onsite Hilbert space dimension, we show that charge transfer fluctuations approach those of the symmetric exclusion process at long times, with subleading $t^{-1/2}$ quantum corrections. 
%In particular, our theory predicts that even in equilibrium, deviations from Gaussianity in current fluctuations are {\em universal} at long times. 
We discuss our results in the context of fluctuating hydrodynamics and macroscopic fluctuation theory of classical non-equilibrium systems, and check our predictions against direct matrix-product state calculations.

\end{abstract}

\maketitle

\textbf{Introduction} - The long-time dynamics of generic many-body quantum systems is expected to be effectively classical. Starting from a pure initial state, the local properties of chaotic systems quickly thermalize: the expectation value of local operators can described by an effective Gibbs ensemble with spatially-varying Lagrange multipliers such as temperature. The resulting evolution from local to global equilibrium is then described by the classical equations of {\em hydrodynamics}. However, the advent of quantum simulator platforms such as cold atoms~\cite{Mazurenko2017,Gross2017,Bakr_2009,FCSSchmiedmayer,Weitenberg_2011,Bohrdt_2021,Parsons_2016,Hilker_2017,Mitra_2017,Haller2015,Sherson2010,bloch2012,Bernien_2017,FCS_KPZ_Bloch}, trapped ions~\cite{Islam_2011,Zhang_2017,Garttner_2017} or superconducting arrays~\cite{Song_2017,GoogleSupremacy,Blais_2021,Wendin_2017} has made it possible to measure not only local expectation values, but also their full quantum statistics. Whether there exists an emergent classical description of such fluctuations in generic, chaotic many-body quantum systems is an open question. 

Consider a one-dimensional quantum system with a conserved charge, that is prepared with a domain-wall chemical potential imbalance across the central bond $\mu_L=-\mu_R=\mu$. By measuring the charge in the right half of the system at times $0$ and $t$, experiments reveal ``quantum snapshots'' of the charge transfer $Q$ across the central bond (from the left to right). By repeating the experiment, one has access the full distribution of measurement outcomes $P_t(Q)$. While the average of that distribution is described by hydrodynamics -- which in the case of a single conserved charge simply reduces to a diffusion equation -- higher cumulants describe current fluctuations and the full counting statistics (FCS) of charge transport~\cite{Levitov1993,PhysRevB.51.4079,Levitov1996ElectronCS,Ivanov1997,Belzig2001,Belzig2002,Levitov_2004,Lesovik13}. 

Computing the FCS in many-body quantum systems is a formidable task, and exact or mean field results have only been achieved in a few cases, notably in non-interacting fermion models~\cite{Najafi2017,Groha2018,QSSEP_Bernard_2021,Bernard_2019,PhysRevB.87.184303,Sch_nhammer_2007,Bernard2022,2022arXiv220411680H}, integrable systems~\cite{XXZFCS,Stephan_2017,Bernard_2016,Bastianello_2018,Myers2018,Calabrese_2020,DeNardis2022,Krajnik2022a,Krajnik2022b,Krajnik2022c,Doyon2022,Bertini2023} and in quantum dots/few qubit models~\cite{Tang_2014,Pilgram_2003,Clerk_2011,FCS_IRLM,Ridley2018,Kilgour2019,Erpenbeck2021,Popovic2021}. While there is currently no exact result pertaining to chaotic many-body quantum systems, charge current fluctuations are expected to be subject to the large deviation principle~\cite{Touchette_2009,Touchette_2013,Lazarescu_2015}: all cumulants of charge transfer should scale in the same way with time, as $\sqrt{t}$ for a diffusive system in one dimension. In the context of classical stochastic models with a conserved charge, the emergence of the large deviation principle is understood within a general formalism known as macroscopic fluctuation theory (MFT)~\cite{Bertini_2015}. MFT is a toolbox for solving the noisy diffusion equation obtained from promoting the hydrodynamic equation to a non-linear {\em fluctuating hydrodynamic} theory by adding a noise term to the current, whose strength are determined by the fluctuation-dissipation theorem. MFT has been very successful in describing stochastic classical systems, and has recently been used to compute the FCS of a paradigmatic integrable Markov chain, the (simple) symmetric exclusion process (SEP)~\cite{Mallick_2022,Dandekar_2022}.

Quantum systems have intrinsic quantum fluctuations, and it is natural to wonder whether they can be captured by  an emergent classical description such as MFT. In this letter, we investigate the FCS in an ensemble of diffusive chaotic models -- random unitary circuits with a conserved $U(1)$ charge~\cite{Rakovszky_2018,Khemaniu1_2018}. Quantum systems with a conserved charge are endowed with current fluctuations and counting statistics. %In random circuits, these fluctuations are controlled by the local Hilbert space dimension $q$, which plays the role of a large-$N$-like perturbative parameter. 
While the quantum many-body dynamics of individual circuit realizations is generally inaccessible, by ensemble averaging, we will study the dynamics of typical circuit realizations. At the level of mean transport, this is known to yield a classical stochastic description \cite{Rakovszky_2018,Khemaniu1_2018,Agrawal_2022,Barratt_2022}. In this work we show that a classical stochastic process in fact describes the entire (late time) FCS, and quantify the sub-leading corrections.

\begin{figure}
  \raisebox{-0.5\height}{\includegraphics[height=0.23\textwidth,width=0.175\textwidth]{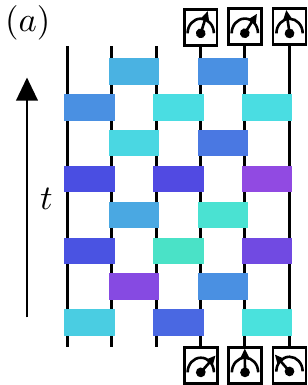}}
  \hspace{-3mm}\raisebox{-0.5\height}{\includegraphics[height=0.268\textwidth]{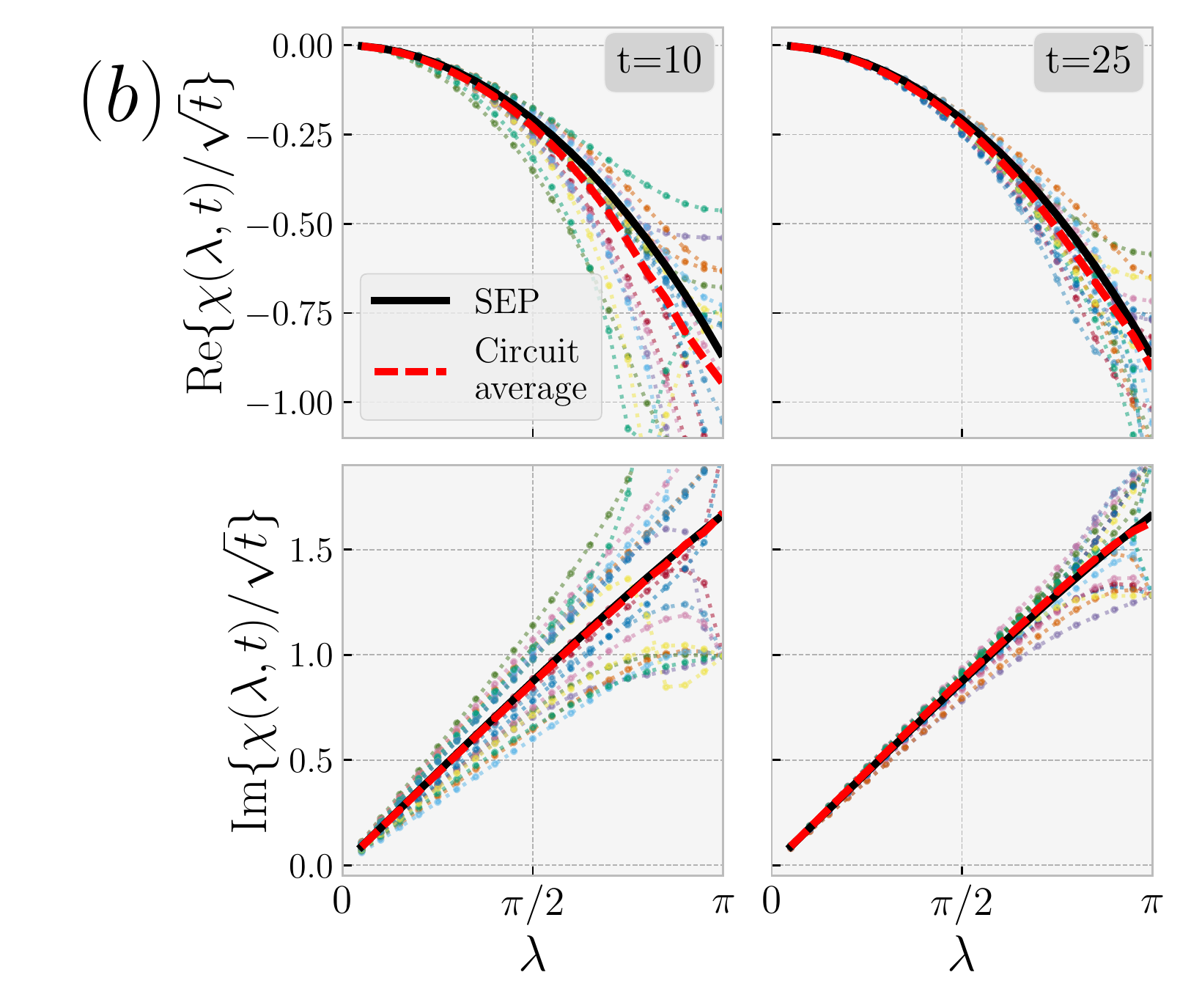}}
  \caption{(a) A two-time measurement protocol for charge transfer across the central bond in a random unitary circuit with a $U(1)$ conserved charge. The charge in the right half of the system is measured at times $0$ and $t$. (b) The cumulant generating function $\chi(\lambda,t)$ with a step initial state ($\mu=\infty$) at times $t=10$ and $t=25$ for different circuit realizations (multi-colored) from TEBD simulations, the circuit averaged CGF with $35$ samples (red dashed) and the late time analytical prediction for the SEP CGF~\cite{Derrida_2009} (black solid). The two FCS snapshots show self-averaging of the FCS; circuit-to-circuit fluctuations in the rescaled CGF $\chi/\sqrt{t}$ decay as $\order{1/t}$~\cite{supp}.}
  
  %The probability distribution $P_t(Q)$ and cumulant generating function $\chi(\lambda)$ for a chaotic diffusive system with diffusivity $D(\rho)=1$ and conductivity $\sigma(\rho)=\rho(1-\rho)$~\cite{Mallick_2022} (corresponding to both the $U(1)$ quantum circuits we consider and SEP) with initial densities $\rho_L=1$, $\rho_R=0$ in the left (L) and right (R) halves of the system.}
  \label{measurement cartoon}
\end{figure}

In order to capture typical current fluctuations within a single circuit realization, circuit averaging must be performed at the level of cumulants, which are polynomial in the system's density matrix. By doing so, we map the problem of computing cumulants onto that of expectation values in replica statistical mechanics (SM) models. By simulating the SM time evolution using matrix-product states, and separately, by introducing an effective stochastic model of coupled SEP chains, we show that the quantum corrections to the higher order cumulants are sub-leading. This leads to a late-time FCS consistent with a simple fluctuating hydrodynamics for xthe coarse grained charge density $\rho(x,\tau)$ ~\cite{Mallick_2022} with re-scaled space-time coordinates $x=j/\ell$ and $\tau=t/\ell^2$,
\begin{equation}\label{fluct-hydro}
\partial_{\tau} \rho  = - \partial_x j , \ j = -D(\rho)\partial_x \rho + \sqrt{\frac{2 \sigma(\rho)}{\ell}}\xi,
\end{equation}
where $\xi(x,\tau)$ is a Gaussian white noise with zero mean and unit variance, and $\ell$ is the size of the hydrodynamic cells over which $\rho$ is coarse-grained. The only microscopic input is this equation are the diffusion constant $D(\rho) =1$ and the conductivity $\sigma(\rho) = D(\rho) \chi_s(\rho)$ with $\chi_s(\rho) = \rho(1-\rho) $, which characterize both random quantum circuits and SEP. The noise term in eq.~\eqref{fluct-hydro} is set by the fluctuation-dissipation theorem to preserve equilibrium charge fluctuations, making this equation a natural candidate for a fluctuating hydrodynamic theory of random quantum circuits. We confirm this result by computing the FCS in individual quantum circuits using matrix product state techniques~\cite{supp} (Fig.~\ref{measurement cartoon}) as an independent check to our effective stochastic theory. Our results establish the emergence of ``classicality'' at long times in quantum systems, even at the level of fluctuations. 

\vspace{1mm}

\textbf{The model and measurement scheme} - We work with a one dimensional chain, in which each site is comprised of a charged qubit with basis states $\ket{q=0,1}$, and a neutral qudit of dimension $d$, yielding a single-site Hilbert space $\mathcal{H}_{\textrm{loc}}\equiv\mathbb{C}^2 \otimes \mathbb{C}^d$. The system evolves via the application of layers of random nearest-neighbor unitary gates in a brick-wall pattern (see Fig. \ref{measurement cartoon}). The unitary gates conserve the total charge on the two sites, but are otherwise Haar random~\cite{Rakovszky_2018,Khemaniu1_2018}. %(Spectral statistics and transport properties have been considered in similar $U(1)$-conserving models \cite{Friedman2019,Singh2021}.)

Unitary evolution and projective measurement ensures that the system's charge dynamics is endowed with current fluctuations. We will investigate the charge transfer $Q$ across the central bond in a time window $[0,t]$ by following the two-time projective measurement protocol~\cite{Kurchan2000,Tasaki2000,Esposito_2009,Campisi_2011,CampisiColloq,Talkner2017} in Fig.~\ref{measurement cartoon}, i.e., measuring the operator $\hat{Q}_R$ for the charge in the right half of the system at times $0$ and $t$. The FCS for this measurement setup is characterized by the cumulant generating function $\chi(\lambda)\equiv \log \langle e^{i\lambda Q} \rangle_t$, where the average $\langle f(Q) \rangle_t = \sum_Q P_t(Q) f(Q)$ is over repetitions of the measurement protocol and $P_t(Q)$ is the probability to measure a charge transfer $Q$. As shown in~\cite{Tang_2014}, writing $P_t(Q)$ in terms of Born probabilities enables us to write the average over measurements as a quantum expectation value~\cite{supp}
\begin{equation}\label{quantum FCS}
\langle e^{i\lambda Q}\rangle_t = \langle \mathcal{T} e^{i\lambda \Delta \hat{Q}_R} \rangle' \equiv \Tr\left[\mathcal{T} e^{i\lambda \Delta \hat{Q}_R}  \hat{\rho}' \right],
\end{equation}
where $\Delta\hat{Q}_R \equiv \hat{Q}_R(t) - \hat{Q}_R(0)$ and $\hat{Q}_R(t)\equiv U(t) \hat{Q}_R U(t)^\dagger$ is the Heisenberg evolved charge operator. The non-commutativity of quantum dynamics requires the use of the time-ordering $\mathcal{T}$~\cite{Levitov1996ElectronCS,Nazarov1999,Nazarov_2003}. The density matrix $\hat{\rho}'$ is related to the initial state $\hat{\rho}$ by the quantum channel $\hat{\rho}'=\sum_q P_q \hat{\rho} P_q$, where $P_q$ are projectors onto the charge sector $Q_R=q$. For initial states with a chemical potential imbalance, $\hat{\rho} \propto \exp[\mu \hat{Q}_L-\mu \hat{Q}_R]$, we simply have $\hat{\rho}'=\hat{\rho}$. 

The circuit averaged charge dynamics is known to maps onto that of a discrete-time symmetric simple exclusion process~\cite{Rowlands_2018,Rakovszky_2018,Khemaniu1_2018} with a brick-wall geometry, i.e., $\overline{P_t(Q)}=P_{t,\textrm{SEP}}(Q)$ where $\overline{O}$ refers to the averaging $O$ over circuits -- all of the quantum fluctuations are lost in the circuit averaged moments of charge transfer. To capture the FCS in typical quantum circuits, we focus on self-averaging quantities, in particular, the cumulants of charge transfer. The cumulants are related to the generating function by $C_m(t) \equiv (-i\partial_\lambda)^m \chi(\lambda) |_{\lambda = 0}$. To compute the $n$-th cumulant, we introduce an often-used $n$-replica statistical mechanics model~\cite{Nahum_2018,Rakovszky_2018,Zhou_2019,Vasseur_2019,Jian_2020,Bao_2020,Potter2022}, expressing each cumulant as a statistical expectation value.
\vspace{1mm}

\begin{figure*}
  \includegraphics[width=0.34\textwidth]{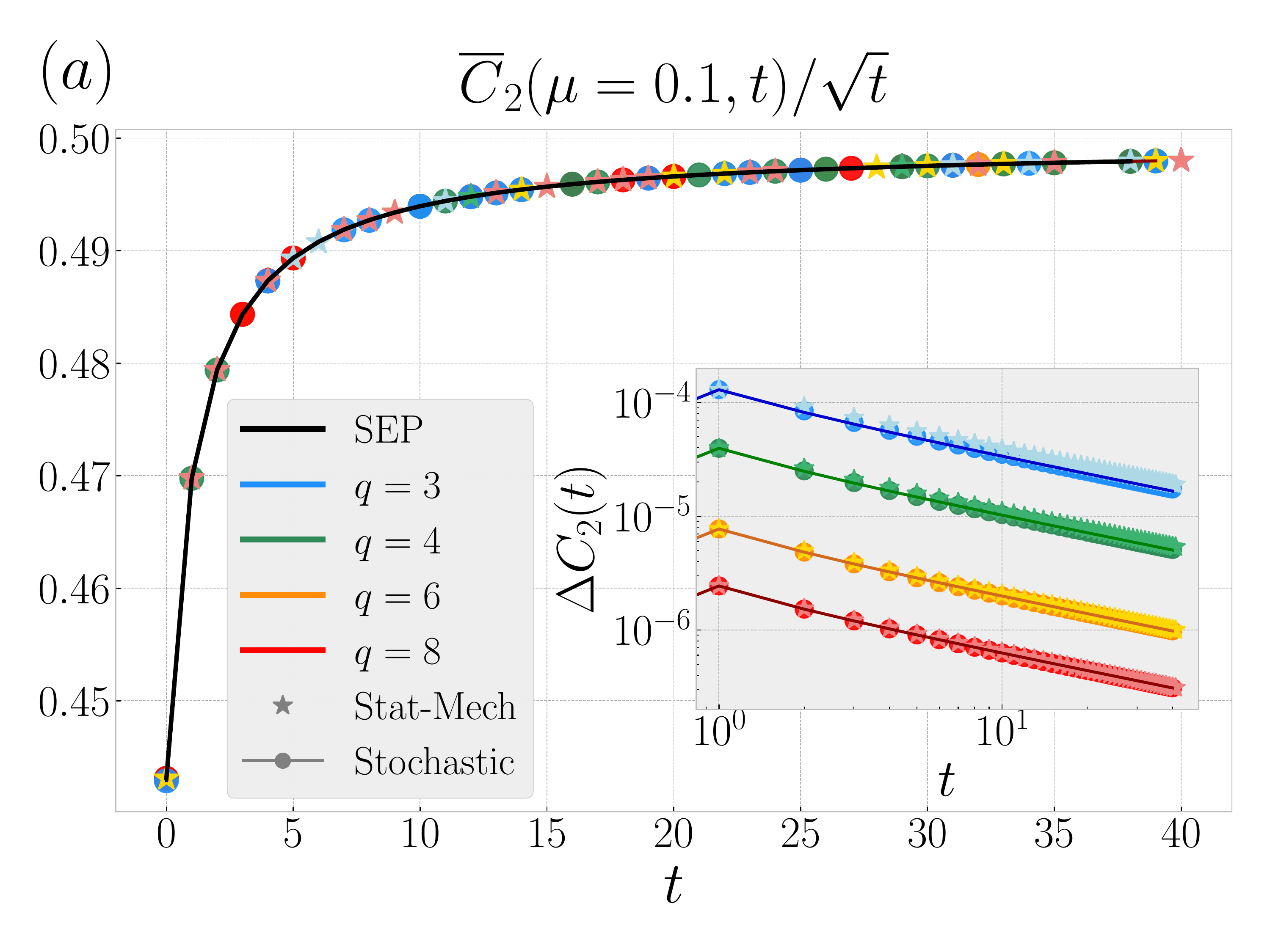}
  \hspace{-3mm}\includegraphics[width=0.34\textwidth]{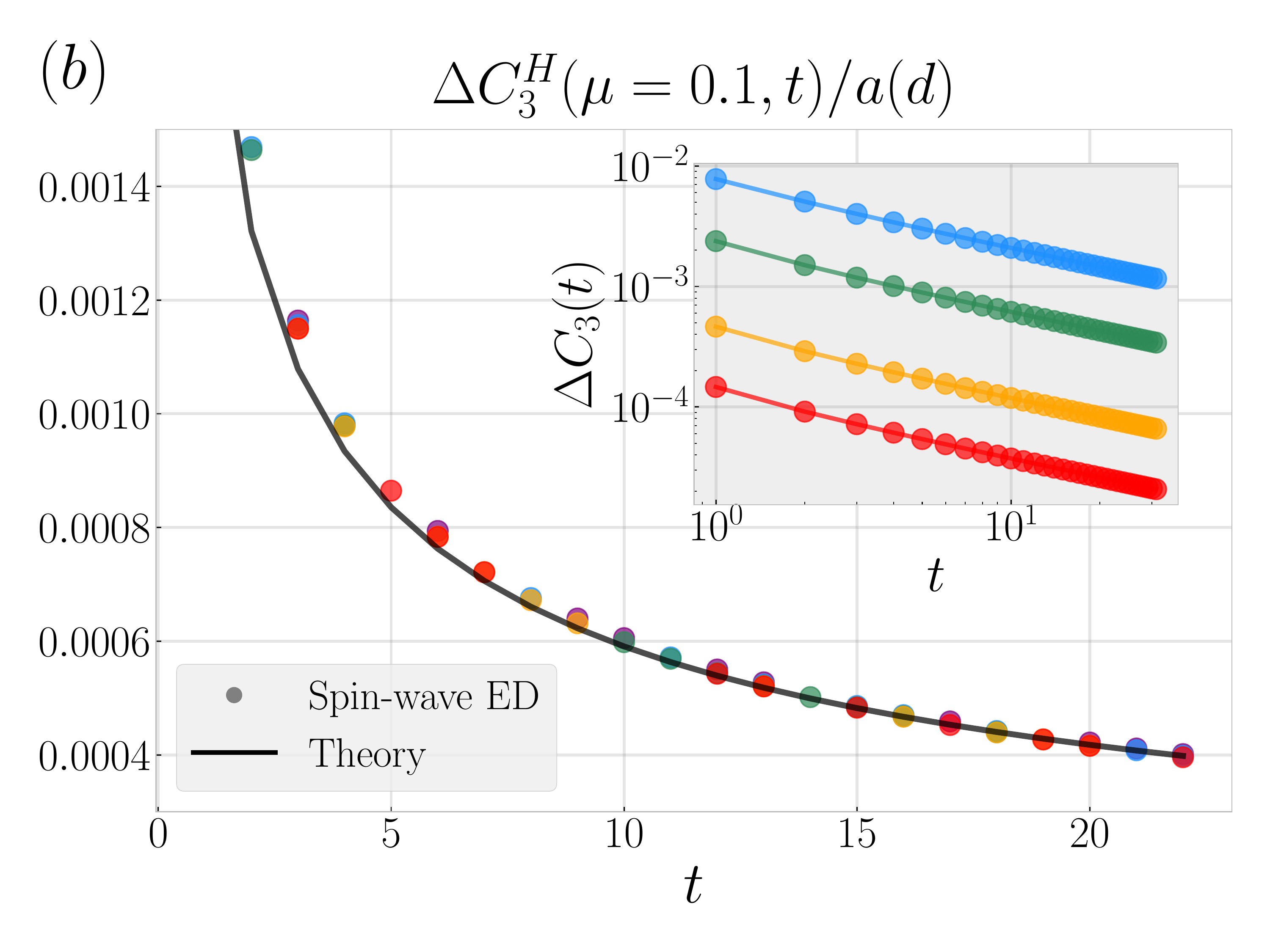}
  \hspace{-3mm}\includegraphics[width=0.34\textwidth]{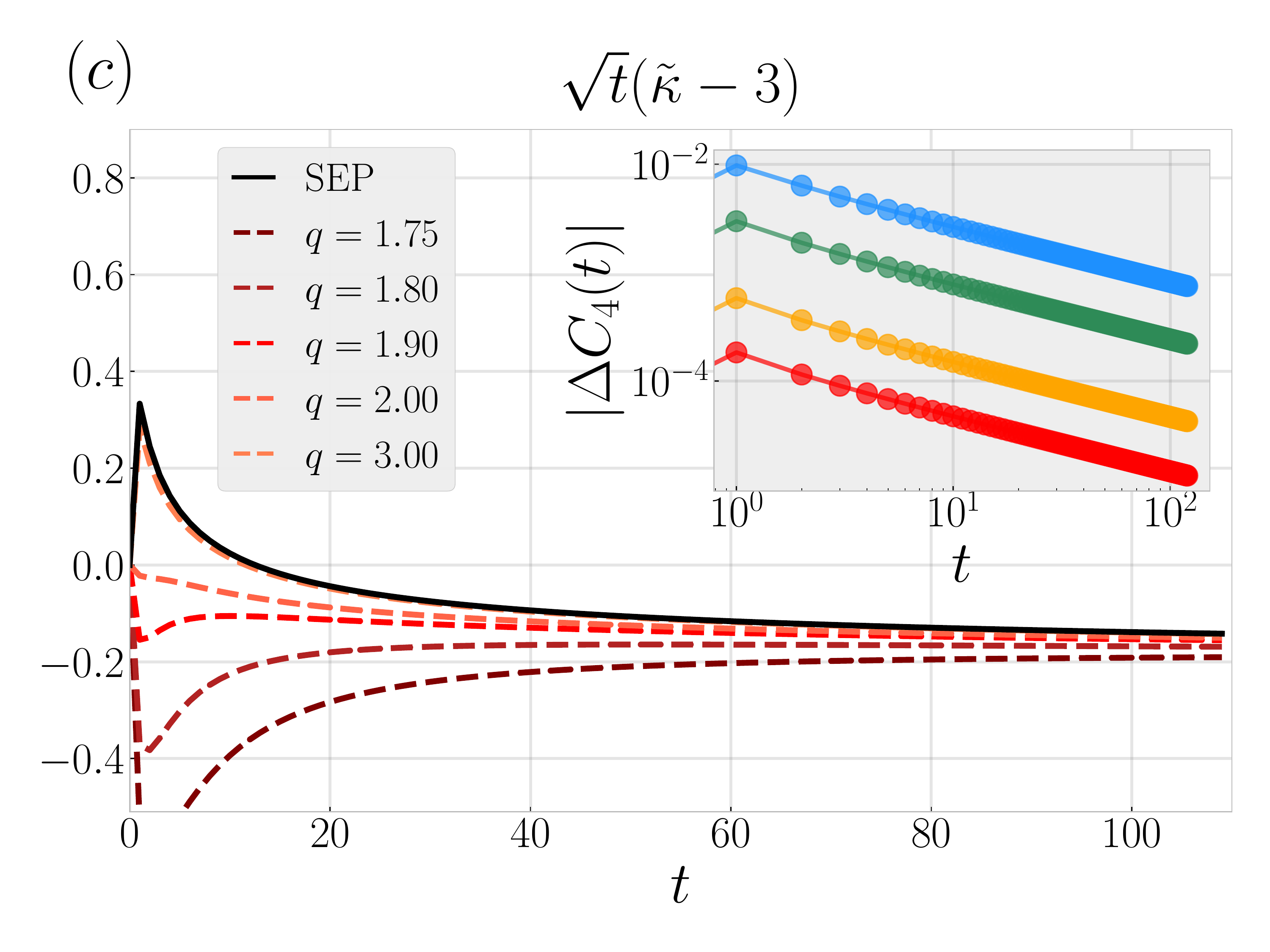}
  \caption{Circuit averaged charge transfer cumulants $\overline{C}_n$ for $U(1)$ charge conserving random unitary circuits at different local Hilbert space dimension $q=3,4,6,8$ and in a discrete-time symmetric simple exclusion process, computed using TEBD applied to the SM transfer matrix: (a) the variance at chemical potential imbalance $\mu=0.1$ (main) and the difference from SEP $\Delta C_2$ (inset) with data from a replica statistical mechanics model and an effective stochastic process; (b) the third cumulant (rescaled by the inter-chain coupling $a(d)$) for a softened stochastic model with Hamiltonian $H_3$ (see eq.~\eqref{Ham}) (main) and the approach to SEP of the circuit averaged third cumulant (inset); (c) a proxy for the excess Kurtosis showing a $t^{-1/2}$ approach to a Gaussian $\kappa=3$ (main), and the approach to SEP of the circuit averaged fourth cumulant at equilibrium (inset).}\label{cumulants}
\end{figure*}

\textbf{Mapping to a statistical mechanics model} - By circuit averaging, we reduce the size of the state space needed to describe the replicated model. The Haar average of a replicated gate, $\overline{\mathcal{U}}\equiv \overline{U^{\otimes n}\otimes U^{*\otimes n}}$, projects onto a smaller space of states characterized by only the local charge degrees of freedom and a permutation degree of freedom $\sigma \in S_n$ that defines a pairing between the $n$ replicas at each site (specifically, between the $n$ conjugated and un-conjugated replicas).

The circuit average of the replicated circuit is equivalent to a statistical mechanics model~\cite{Nahum_2018,Rakovszky_2018,Zhou_2019,Vasseur_2019,Jian_2020,Bao_2020,Li2021,Fisher2022,Dias2022,Friedman2019,Singh2021,Agrawal_2022} with the permutation degrees of freedom living on the vertices and the charge configurations on the edges. The partition function for this statistical mechanics model is given by a sum over the charge configurations and permutations (compatible with the charges) with statistical weights associated with each edge~\cite{Rakovszky_2018,Agrawal_2022,Barratt_2022}. 

In the SM model, $d\to\infty$ locks together neighboring permutations, and together with the initial and final boundary conditions $\sigma_0=\sigma_t=\mathbb{1}$, the $n$-replica model decouples into $n$ independent discrete-time SEP chains. Letting $d$ be large but finite allows different permutations to appear during the dynamics; domain walls between domains of different permutations $\sigma$ and $\tau$ have an energy cost of $\order{|\sigma\tau^{-1}|\log(d)}$ per unit length of domain wall~\cite{Zhou_2019} ($|\sigma|$ is the transposition distance of $\sigma$ from $\mathbb{1}$). This is the basis of a large-$d$ expansion that is the focus of the next section.

We use the time-evolving block decimation (TEBD) algorithm~\cite{TEBD,TEBD2,SCHOLLWOCK201196} to apply the $n=2$ SM transfer matrix, and compute exactly the charge transfer variance, $\overline{C}_2$, which is given as a SM expectation value. Denoting the $n$-replica expectation value by $\langle \cdot \rangle_{n\textrm{-rep}}$, and using superscripts to indicate in which replica an observable acts, the variance is given by % second and third cumulants are given by
\begin{equation}\label{SM exp}
\overline{C_2}(t) = \langle \mathcal{T}\Delta\hat{Q}^{(1) 2}_R-\Delta\hat{Q}^{(1)}_R \Delta\hat{Q}^{(2)}_R \rangle_{2-\textrm{rep}}.
\end{equation}
%\begin{align}\label{SM exp}
%\overline{C_2}(t) = \langle &\mathcal{T}\Delta\hat{Q}^{(1) 2}_R-\Delta\hat{Q}^{(1)}_R \Delta\hat{Q}^{(2)}_R \rangle_{2-\textrm{rep}}, \\
%\overline{C_3}(t) = \langle &\mathcal{T}\Delta\hat{Q}^{(1) 3}_R-3\mathcal{T}\Delta\hat{Q}^{(1) 2}_R \Delta\hat{Q}^{(2)}_R \nonumber \\
%&+2\Delta\hat{Q}^{(1)}_R \Delta\hat{Q}^{(2)}_R  \Delta\hat{Q}^{(3)}_R  \rangle_{3-\textrm{rep}}.
%\end{align}
Using maximum bond dimension $\chi = 1500$, we compute $\overline{C}_2$ for different initial chemical potential imbalances $\mu$ and for local Hilbert space dimensions $q \equiv 2d = 3,4,6,8$ \footnote{This selection of local Hilbert space dimensions corresponds to qudit dimensions $d=1.5,2,3,4$. In the SM model, $d$ is just a parameter and need not be physical (integer).}. The results for $\mu=0.1$ are shown in first panel of Fig. \ref{cumulants} and results for $\mu=2$ and $\infty$ can be found in the supplementary materials~\cite{supp}. By subtracting the variance for $q=\infty$ (i.e., the SEP variance), we isolate the quantum contributions to $\overline{C}_2$, which we call $\Delta C_2$, and find that these decay as $t^{-1/2}$ for all $q$ (inset of panel 1, Fig. \ref{cumulants}). The $n$-replica SM model requires a local state space of dimension $2^n n!$, putting higher cumulants beyond reach with TEBD. In order to access the higher cumulants, and to find a theoretical explanation for the approach to SEP at $n=2$, we develop an effective stochastic model for the charge dynamics in the replicated SM models. 

%The results for the variance are shown in first panel of Fig. \ref{cumulants} for an initial chemical potential imbalance of $\mu=0.1$; the inset shows a $t^{-1/2}$ approach to the purely SEP case ($d\to\infty$).

\vspace{1mm}

\textbf{An effective stochastic model} - At large $d$, the lowest energy contributions to the SM free energy come from dilute configurations of small domains of single transpositions in an `all-identity' background. The smallest of these domains -- or \textit{bubbles} -- have the lowest possible energy cost of $4\log(d)$. All configurations of these bubbles can be counted in the brick-wall circuit picture by inserting a projector $P_\mathbb{1}$ onto the identity permutation sub-space in-between every replicated gate $\overline{\mathcal{U}}$. 

Upon doing this, we can replace $\overline{\mathcal{U}}$ with a gate $G_{(n)}$ that explores only the $\sigma=\mathbb{1}$ subspace but has a modified charge dynamics~\cite{supp},
\begin{equation}
\includegraphics[width=0.1\textwidth, valign=c]{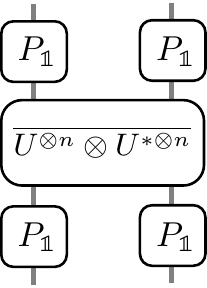}\ =\ \includegraphics[width=0.1\textwidth, valign=c]{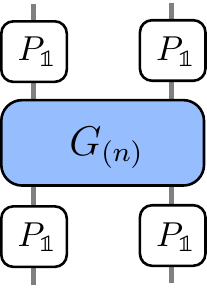} \ .
\end{equation}
The result is an effective Markov process described by an $n$-chain ladder with hard-core random walkers on each chain and a hopping rate that is conditional on the local occupancy of the other chains. More concretely, the model is that of $n$ discrete-time SEP chains with pairwise local interactions between chains -- when two chains have the same (different) charge configuration at a pair of neighboring sites, the interaction biases transitions in favor of states in which both chains have the same (different) configurations. The transfer matrix is given by a product of even and odd layers of two-site operators, $T=T_{E}T_{O}$ with $T_{E/O} = \prod_{j \in \textrm{Even}/\textrm{Odd}} T_{j,j+1}$. Representing a charge with a red dot and focusing on $n=2$ replicas (labelled $1,2$), the modified transitions on a pair of sites $(i,j)$ are given by
\begin{align}
    \raisebox{-0.41\totalheight}{\includegraphics[width=0.93cm]{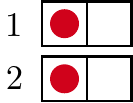}}\ \raisebox{-0.3cm}{,} \ \raisebox{-0.42\totalheight}{\includegraphics[width=0.65cm]{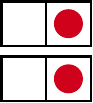}} &\to p\ \raisebox{-0.42\totalheight}{\includegraphics[width=0.65cm]{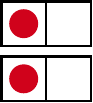}} + p\ \raisebox{-0.42\totalheight}{\includegraphics[width=0.65cm]{TransferMatFig4.pdf}}
    +r\ \raisebox{-0.42\totalheight}{\includegraphics[width=0.65cm]{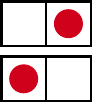}} +r\ \raisebox{-0.42\totalheight}{\includegraphics[width=0.65cm]{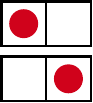}} \nonumber \\
    \raisebox{-0.58\totalheight}{\includegraphics[width=0.93cm]{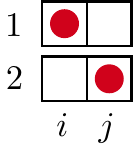}}\ \raisebox{-0.3cm}{,} \ \raisebox{-0.58\totalheight}{\includegraphics[width=0.65cm]{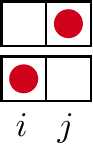}} &\to r\ \raisebox{-0.42\totalheight}{\includegraphics[width=0.65cm]{TransferMatFig1.pdf}} + r\ \raisebox{-0.42\totalheight}{\includegraphics[width=0.65cm]{TransferMatFig4.pdf}}
    +p\ \raisebox{-0.42\totalheight}{\includegraphics[width=0.65cm]{TransferMatFig2.pdf}}
    +p\ \raisebox{-0.42\totalheight}{\includegraphics[width=0.65cm]{TransferMatFig3.pdf}}\ ,
\end{align}
where the transition probabilities are $p=\frac{1+a}{4}$ and $r=\frac{1-a}{4}$ with $a(d)=[4d^4-1]^{-1}$. All other transitions are as given for decoupled SEP chains (charges hopping with probability $1/2$). The derivation of the Markov process is described in detail in the supplementary materials~\cite{supp}. 

This effective model inherits an $n$-fold $SU(2)$ invariance (one for each chain) from the SM model, allowing for arbitrary rotations of the charge basis $\left|Q=0,1\right)$ in each chain (see supplementary materials for details). Choosing a rotated basis ($\{\left|\uparrow\right) \propto \left|0\right)+\left|1\right), \left|\downarrow\right) \propto \left|0\right)-\left|1\right)\}$), the $n$-th cumulant can be written in terms of matrix elements of the $n$-chain transfer matrix, $T_n$, with the initial and final states having at most $n$ magnons (overturned spins). This reduces the problem of calculating $C_n$ to the diagonalization of an $L^n\times L^n$ matrix. 
%(this is shown for $n=2$ in the supplementary materials, where $C_2$ is computed by means of a spin-wave calculation).
\vspace{1mm}

\textbf{Results} -  By applying the Markov process transfer matrix exactly, we calculate the second and third cumulants at different biases and the fourth cumulant in equilibirum. We find that in all cases, the effective evolution approaches SEP as $\Delta C_n \equiv \overline{C}_n-C^{\textrm{SEP}}_n \sim a(d)t^{-1/2}$ %where $a(d)=[4 d^4 -1 ]^{-1}$ 
(see the insets in Fig. \ref{cumulants} and~\cite{supp}). The variance data shows excellent agreement between the SM model and the effective model. 
%Cumulant data for chemical potential imbalance $\mu=2,\infty$ can be found in the supplementary materials.

In chaotic models at equilibrium (no bias, $\mu=0$), we expect that the distribution $P_t(Q)$ will approach a Gaussian at late times. However, even long-time deviations from Gaussianity are {\em universal} and are captured by an effective classical stochastic  model -- SEP in the case of random circuits. For example, using standard SEP results~\cite{Derrida_2009}, we find that at half-filling, the average equilibrium excess Kurtosis decays in a universal way as
\begin{equation}
\kappa - 3 = \frac{(4- 3 \sqrt{2})  \sqrt{\pi}}{2\sqrt{t}} + \dots
\end{equation}
independently of the value of $q$. Circuit averaging quantities with the evolution unitary in the denominator, such as Kurtosis, requires a replica trick. To avoid this, we calculate the proxy $\tilde{\kappa}\equiv\frac{\overline{\mu_4}}{\overline{\sigma^4}}$ that averages the numerator and denominator separately ($\mu_4$ is the fourth central moment and $\sigma$ is the standard deviation) and find the same universal approach to a Gaussian, $\kappa = 3$, for different $q$ (panel 3 of Fig. \ref{cumulants}). We have accentuated the variations between models by using unphysical local Hilbert space dimensions $q$.
\vspace{1mm}

{\bf Effective Hamiltonian} - To understand the approach to SEP at long times, we can map the effective $n$-chain Markov processes to an effective ferromagnetic Hamiltonian. We do this by softening the transfer matrix, $T_n \to e^{-H_n}$. The effective Hamiltonian is given by
\begin{equation}\label{Ham}
H_n \equiv \sum_j \sum_{\alpha=1}^n P^{(\alpha)}_{j,j+1} - a(d) \sum_j  \sum_{ \alpha < \beta } P^{(\alpha)}_{j,j+1} P^{(\beta)}_{j,j+1},
\end{equation}
where the superscripts indicate in which chain an operator acts and where the second term contains a sum over distinct pairs of chains. We have dropped sub-leading $\order{1/d^8}$ terms. In terms of Heisenberg spin interactions, the projector $P$ is given by $P_{j,j+1} = \frac{1}{4} - \boldsymbol{S}_j\cdot \boldsymbol{S}_{j+1}$. The imaginary time dynamics is then dominated at late times by the low energy physics of~\eqref{Ham}. We study the low energy spectrum for $n=2$ using standard spin-wave methods~\cite{supp} and find that, at late times, the quantum contribution to the charge transfer variance is
\begin{equation}
\Delta C^H_2 \approx \frac{a \tanh(\mu/2)^2}{16\sqrt{\pi t}},
\end{equation}
where the superscript $H$ indicates that this prediction is for the continuous time stochastic model with imaginary time Hamiltonian dynamics~\cite{supp}. We also consider the third cumulant in the softened stochastic model, finding the familiar $t^{-1/2}$ decay of quantum fluctuations (Fig. \ref{cumulants} panel 2) from numerics and theoretical predictions in the linear response regime ($\mu\ll 1$~\cite{supp}). This general scaling can be generalized to higher cumulants using a simple renormalization group (RG) argument based on power-counting: because of the imaginary time evolution, the long-time dynamics is controlled by the low energy-properties of eq.~\eqref{Ham}. Using standard spin-coherent state path integral techniques, it is straightforward to show that the perturbation coupling the replicas with strength $a(d)$ has scaling dimension $\Delta=4$, and is thus irrelevant in the RG sense. At long-times, we thus expect the different replicas (SEP chains) to be effectively decoupled so that $\langle O \rangle_{n-\textrm{chain}} = \langle O \rangle_{\textrm{SEP}}(1+\order{t^{-1}})$. % where we have used the $z=2$ (diffusive) dynamics of the unperturbed Hamiltonian. 

The asymptotic decoupling between replicas also establishes that circuit-to-circuit fluctuations are suppressed at long times. To see this, consider an $n$-copy quantity $A$ (this could be mean charge transfer for $n=1$ or charge transfer variance for $n=2$), the circuit average of $A$ is given by $\overline{A}=\langle X \rangle_{n\textrm{-chain}}$ for some operator $X$ on $n$ replicas, whereas the circuit-to-circuit fluctuations is controlled by $\overline{(A-\overline{A})^2} = \langle X\otimes X \rangle_{2n\textrm{-chain}}-\langle X \rangle_{n\textrm{-chain}}^2$. Using the asymptotic decoupling of the SEP chains we have the aforementioned suppression of circuit-to-circuit fluctuations, $\textrm{Var}(A)/\overline{A}^2 \sim t^{-1}$. Therefore, the FCS of individual quantum circuits approaches the SEP predictions as
\begin{equation}
\chi(\lambda)/\sqrt{t} = \overline{\chi(\lambda)}/\sqrt{t} + {\cal O}(1/t),
\end{equation}
with $\overline{\chi}/\sqrt{t} \to \chi_{\rm SEP}/\sqrt{t}$ as $t \to \infty$.
%Our results are thus expected to apply to individual realizations of random quantum circuits, and more broadly to all chaotic many-body quantum systems. 
To verify this prediction, we have computed the FCS of individual random quantum circuits for a domain wall initial state ($\mu=\infty$) using standard counting field techniques~\cite{supp} (Fig.~\ref{measurement cartoon}). We find that the rescaled CGF $\chi(\lambda)/\sqrt{t}$ is indeed self-averaging with  $\mathcal{O}(1/t)$ fluctuations, and does approach the SEP predictions at long times~\cite{supp}. 

%Circuit-to-circuit fluctuations are discussed further in the supplemental material~\cite{supp}.
\vspace{1mm}

\textbf{Discussion} - 
Our main result is that charge transfer fluctuations in random charge-conserving quantum circuits is controlled by an effective SEP stochastic model at long times: $C_n = C^{\textrm{SEP}}_n +\order{t^{-{1/2}}}$. The full cumulant generating function of individual random circuits $\chi(\lambda)\approx \overline{\chi (\lambda)} $ must then take the same form as that of SEP at late times, $\overline{\chi (\lambda)} \equiv \overline{\log \langle e^{i \lambda Q}\rangle } \approx \chi_{\textrm{SEP}}(\lambda)$. The symmetric exclusion process generating function is known analytically~\cite{Derrida_2009} from integrability, and is given by
\begin{equation}\label{Derrida}
\chi(\lambda) \approx \sqrt{t} F(\omega), \ F(\omega)=\frac{1}{\sqrt{\pi}}\sum_{n = 1}^\infty \frac{(-1)^{n+1}}{n^{3/2}}\omega^n,
\end{equation}
where $\omega = \rho_L(e^{i\lambda}-1) + \rho_R(e^{-i\lambda}-1) + \rho_L \rho_R(e^{i\lambda}-1)(e^{-i\lambda}-1)$ and $\rho_{L/R}=\frac{e^{\mu_{L/R}}}{1+e^{\mu_{L/R}}}$ is the initially local charge density in the left ($L$) and right ($R$) halves of the system~\footnote{This result is for a continuous time SEP rather than the discrete time variant. However, since both share the same diffusion constant $D(\rho)=1$ and conductivity $\sigma(\rho)=\rho(1-\rho)$, they share the same FCS~\cite{Bertini_2015}.}. %Remarkably, our results relate this exact, fine-tuned result relying on the integrability of the SEP model to the {\em generic} behavior of chaotic many-body quantum systems.  
The same FCS was recently shown to emerge from MFT~\cite{Mallick_2022} from solving eq.~\eqref{fluct-hydro} directly.
Our results thus establish that the current fluctuations of {\em individual realizations} of random quantum circuits are described by the simple fluctuating hydrodynamic equation~\eqref{fluct-hydro}. To fully establish the validity of MFT to many-body quantum systems, it would be interesting to consider ensembles of circuits with more general diffusion constants $D(\rho)$: there as well we expect a similar mapping onto effective classical stochastic models to the one we have found here, with irrelevant inter-replica couplings as in~\eqref{Ham}. We leave the study of such generalizations to future work. 

{\bf Acknowledgements} - 
 We thank Immanuel Bloch, Enej Ilievski, Vedika Khemani, Ziga Krajnik,  Alan Morningstar, Andrew Potter, Tomaz Prosen, and Andrea De Luca for helpful discussions. This work was supported by the ERC Starting Grant 101042293 (HEPIQ) (J.D.N.), the National Science Foundation under NSF Grants No. DMR-1653271 (S.G.) and DMR-2104141 (E.M.),  the US Department of Energy, Office of Science, Basic Energy Sciences, under Early Career Award No.~DE-SC0019168 (R.V.), and the Alfred P. Sloan Foundation through a Sloan Research Fellowship (R.V.).

\bibliographystyle{apsrev4-1} % Tell bibtex which bibliography style to use
\bibliography{referencesv2} % Tell bibtex which .bib file to use (this one is some example file in TexLive's file tree)

\pagebreak

\widetext

\newpage

\makeatletter
\begin{center}
\textbf{\large Supplemental Materials: Full Counting Statistics of Charge in Chaotic Many-body Quantum Systems}

\vspace{5mm}

Ewan McCulloch,\textsuperscript{1,2} Jacopo De Nardis,\textsuperscript{3} Sarang Gopalakrishnan, \textsuperscript{2} Romain Vasseur\textsuperscript{1}

\vspace{1mm}

\textsuperscript{1}\textit{\small Department of Physics, University of Massachusetts, Amherst, MA 01003, USA}

\textsuperscript{2}\textit{\small Department of Electrical and Computer Engineering,\\
Princeton University, Princeton, NJ 08544, USA}

\textsuperscript{3}\textit{\small Laboratoire de Physique Th\'eorique et Mod\'elisation, CNRS UMR 8089,\\
CY Cergy Paris Universit\'e, 95302 Cergy-Pontoise Cedex, France}

\makeatother

\end{center}
%%%%%%%%%% Merge with supplemental materials %%%%%%%%%%
%%%%%%%%%% Prefix a "S" to all equations, figures, tables and reset the counter %%%%%%%%%%
\setcounter{equation}{0}
\setcounter{figure}{0}
\setcounter{table}{0}
\setcounter{page}{1}
\makeatletter
\renewcommand{\theequation}{S\arabic{equation}}
\renewcommand{\thefigure}{S\arabic{figure}}
%\renewcommand{\bibnumfmt}[1]{[S#1]}
%\renewcommand{\citenumfont}[1]{S#1}
%%%%%%%%%% Prefix a "S" to all equations, figures, tables and reset the counter %%%%%%%%%%

\section{Full Counting Statistics in Quantum Mechanical Systems}
%Full counting statistics was initially introduced as a characterization of fluctuations in classical stochastic systems~\cite{Pilgram_2003,Roche_2005,Bertini_2006,Derrida_2009,}.
Full counting statistic was originally introduced as a characterization of current fluctuations in mesoscopic conductors~\cite{Levitov1993,Levitov1996ElectronCS,Ivanov1997,Belzig2001,Belzig2002,Levitov_2004}. In this appendix, we discuss the generalization to generic quantum mechanical systems. In particular, we will closely follow Ref.~\cite{Tang_2014} and define the FCS using a two-time projective measurement protocol. In the two-time measurement protocol, we first measure a quantity of interest -- for example, the charge $q$ in right-hand side of the system -- at time $0$, and then again at time $t$. With the probability of these measurement outcomes denoted $P(q_0,q_t)$, the moment generating function $Z(t,\lambda)$ is defined as the Fourier transform of the probability $P_t(Q) = \sum_{q_0, q_t}P(q_0,q_t)\delta_{q_t-q_0, Q}$ of charge transfer $Q$ (from left to right) in a time window $[0,t]$.
\begin{equation}
Z(t,\lambda) = \sum_{Q} e^{i\lambda Q} P_t(Q).
\end{equation}
From this generating function we also have access to the cumulant generating function $\chi(t,\lambda)$ using the following relation,
\begin{equation}
\chi(t,\lambda) \equiv \log( Z(t,\lambda)).
\end{equation}
The cumulants $C_m$ are then computed as derivatives of $\chi(t,\lambda)$ with respect to the counting field $\lambda$,
\begin{equation}
C_m(t) \equiv (-i\partial_{\lambda})^m\chi(t,\lambda)|_{\lambda=0}.
\end{equation}
The first few cumulants are given by,
\begin{equation}
C_1(t) = \langle Q \rangle_t,\quad
C_2(t) = \langle (Q - \langle Q \rangle_t)^2 \rangle_t,\quad 
C_3(t) = \langle (Q - \langle Q \rangle_t)^3 \rangle_t,
\end{equation}
where $\langle f(Q) \rangle_t \equiv \sum_{Q} P_t(Q) f(Q)$.

We will now relate the generating functions to quantum expectation values. To do this, we implement the projective measurements at time $0$ and $t$ with the projectors $P_{q_0}$ and $P_{q_t}$ respectively. Then, for a given initial state $\ket{\psi}$ at time $0$, the Born probability for the measurement outcomes $q_0$ and $q_t$ is then given by
\begin{equation}
P(q_0,q_t) = \bra{\psi}P_{q_0}U(0,t)P_{q_t}U(t,0)P_{q_0}\ket{\psi},
\end{equation}
with $U$ the unitary evolution operator.  
Generalizing to a mixed initial state $\rho$, $P(q_0,q_t)$ is given by
\begin{equation}
P(q_0,q_t) = \Tr\left[ \rho P_{q_0}U(0,t)P_{q_t}U(t,0)P_{q_0} \right].
\end{equation}
The generating function $Z(t,\lambda)$ is then given by
\begin{equation}
Z(t,\lambda) = \sum_{q_0,q_t} e^{i\lambda (q_t-q_0)} \Tr\left[ \rho P_{q_0}U(0,t)P_{q_t}U(t,0)P_{q_0} \right].
\end{equation}
Denoting $\hat{Q}_R$ as charge operator for the right-hand half of the system, the generating function can be recast as
\begin{align}
Z(t,\lambda) &= \sum_{q_0,q_t} \Tr\left[ \rho P_{q_0} e^{-i\lambda \hat{Q}_R} U(0,t)P_{q_t} e^{i\lambda \hat{Q}_R} U(t,0)P_{q_0} \right], \notag \\
 &= \Tr\left[ \rho' e^{-i\lambda \hat{Q}_R} U(0,t) e^{i\lambda \hat{Q}_R} U(t,0) \right],
\end{align}
where we have used completeness $\sum_{q}P_{q}=\mathbb{1}$ and defined $\rho' = \sum_{q}P_q\rho P_q$. Notice that $\rho'$ remains a valid mixed state as $\sum_{q}P_q\rho P_q$ is a completely positive trace preserving map. Using the notation $\langle \cdot \rangle' = \Tr\left[\rho' \cdot \right]$ and defining the Heisenberg evolved operators $\hat{Q}_R(t)=U(0,t) \hat{Q}_R U(t,0)$, we have the more compact expression for $Z(t,\lambda)$,
\begin{equation}
Z(t,\lambda) = \langle \mathcal{T} e^{i\lambda \Delta \hat{Q}_R (t)}  \rangle',
\end{equation}
where $\Delta \hat{Q}_R (t) = \hat{Q}_R(t) - \hat{Q}_R(0)$ and where $\mathcal{T}$ is the time ordering operator. The cumulants can now be expressed as quantum expectation values,
\begin{equation}
C_1(t) = \langle \Delta \hat{Q}_R(t) \rangle',\quad
C_2(t) = \langle \mathcal{T}(\Delta \hat{Q}_R(t)-\langle \Delta \hat{Q}_R(t) \rangle')^2 \rangle',\quad
C_3(t) =  \langle \mathcal{T} (\Delta \hat{Q}_R(t)-\langle \Delta \hat{Q}_R(t) \rangle')^3 \rangle'.
\end{equation}

\section{Mapping to a statistical mechanics model}
When circuit averaging quantities that are polynomial in evolution unitary of a system, such as entanglement entropies, it is necessary to use a replicated Hilbert space, $\mathcal{H}^{\otimes n}\otimes \mathcal{H}^{*\otimes n}$. Circuit averaging then naturally maps random unitary circuit models onto replica statistical mechanics models~\cite{Nahum_2018,Rakovszky_2018,Zhou_2019,Vasseur_2019,Bao_2020,Li2021,Fisher2022,Dias2022}, whose local degrees of freedom at each site are the local charges $q_m\in\{0,1\}$, $m = 1,\cdots,n$, and permutation degrees of freedom $\sigma \in S_n$. The permutation degrees of freedom define a pairing between the $n$ copies of $\mathcal{H}_{\textrm{loc}}$ and the $n$ copies of $\mathcal{H}^*_{\textrm{loc}}$, where $\mathcal{H}_{\textrm{loc}}$ is the single site Hilbert space in the underlying circuit. In terms of the single site basis states of the quantum circuit, the SM local states are given by
\begin{equation}
\ket{q_1,\cdots,q_n ;\sigma} \equiv \sum_{\substack{a_1,\cdots, a_n\\ a_m \in\mathcal{H}_{q_m}}}\ket{a_1 \cdots a_n a^*_{\sigma(1)} \cdots a^*_{\sigma(n)}},
\end{equation}
where $\mathcal{H}_q\subset\mathcal{H}_{\textrm{loc}}$ is the local charge sector with charge $q$ and the notation $a^*$ distinguishes the states that live in the complex conjugate replicas $\mathcal{H}_{\textrm{loc}}^*$. The central object in the computation of the SM transfer matrices is the circuit average of replicated two-site gates, $\overline{U^{\otimes n}\otimes U^{*\otimes n}}$, which takes the generic form,
\begin{equation}
\overline{U^{\otimes n}\otimes U^{*\otimes n}} = \sum_{\sigma,\tau \in S_n} \sum_{Q_1,\cdots Q_n} W_{\sigma,\tau}(\{Q_m\}) \ket{\{Q_m\};\sigma}\bra{\{Q_m\};\tau},
\end{equation}
where $Q_m=q_{m}+q'_{m}$ is the total charge on both sites in the $m$-th replica, and the states $\ket{\{Q_m\};\sigma}$ are given in terms of the local SM states by
\begin{equation}
\ket{\{Q_m\};\sigma} = \sum_{\substack{q_{1}\cdots q_{n} \\ q'_{1}\cdots q'_{n}}}\Big(\prod_{m}\delta_{q_{m}+q'_{m},Q_m} \Big) \ket{\{q_{m}\};\sigma}\otimes\ket{\{q'_{m}\};\sigma}.
\end{equation}
The functions $W_{\sigma,\tau}(\{Q_m\})$ are computed explicitly for $n=2$ in Ref.~\cite{Rakovszky_2018} and are given by
\begin{equation}
W_{\mathbb{1},\mathbb{1}} = W_{\boldsymbol{\times},\boldsymbol{\times}} =  [d_{Q_1}d_{Q_2}-\delta_{Q_1,Q_2}]^{-1}, \quad W_{\mathbb{1},\boldsymbol{\times}} = W_{\boldsymbol{\times},\mathbb{1}} =  -\delta_{Q_1,Q_2}[d_{Q_1}(d^2_{Q_1}-1)]^{-1},
\end{equation}
where $d_{0}=d_{2}=d^2$ and $d_1=2d^2$ and where $\mathbb{1}$ and $\boldsymbol{\times}$ are the identity and swap permutation respectively. 

\begin{figure}
    \centering
    \hspace{1cm}\raisebox{-0.572\height}{\includegraphics[height=2.875cm]{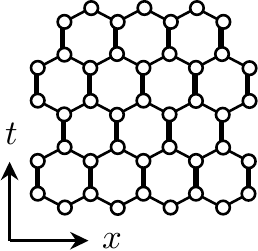}} \hspace{1cm} \raisebox{-0.5\height}{\includegraphics[height=2.5cm]{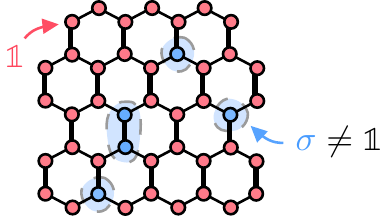}}
    \caption{(Left) The statistical mechanics models generated by circuit averaging replicated circuits live on an an-isotropic hexagonal lattice with the edges carrying charge degrees of freedom and the vertices carrying permutations degrees of freedom. (Right) Dilute configurations of small domains of non-identity permutations dominate the statistical mechanics partition function in the large $d$ limit.}
    \label{hex_lattice}
\end{figure}

The general $n$-replica case is considered in Ref.~\cite{Agrawal_2022}, which we summarize now. Letting $\boldsymbol{q}\equiv (q_1,\cdots,q_n)$ and $\boldsymbol{q}'\equiv(q'_1,\cdots,q'_n)$ be vectors of the charges in each replica on the incoming legs of the two sites and $\boldsymbol{p}$ and $\boldsymbol{p}'$ be the charges on the outgoing legs, we may write $\overline{U^{\otimes n}\otimes U^{*\otimes n}}$ as a tensor diagram,
\begin{equation}\label{generalgate}
\overline{U^{\otimes n}\otimes U^{*\otimes n}} = \sum_{\sigma,\tau\in S_n}\sum_{\boldsymbol{q},\boldsymbol{q}',\boldsymbol{p},\boldsymbol{p}'} \delta(\sigma\tau^{-1}\in\textrm{Stab}(\boldsymbol{q}+\boldsymbol{q}')) \delta_{\boldsymbol{q}+\boldsymbol{q}',\boldsymbol{p}+\boldsymbol{p}'}\textrm{Wg}(\sigma\tau^{-1})
\includegraphics[width=0.1\textwidth, valign=c]{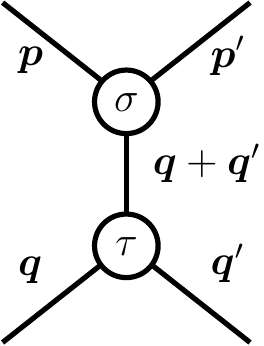},
\end{equation}
where $\textrm{Stab}(\boldsymbol{q}+\boldsymbol{q}')=S_{n_0}\times S_{n_1}\times S_{n_2}$ is the stabilizer group for the configuration of charges $\boldsymbol{Q}=\boldsymbol{q}+\boldsymbol{q}'$, with $n_{Q}$ counting the number of times the charge $Q$ appears in $\boldsymbol{Q}=(Q_1,\cdots,Q_n)$. The function $\textrm{Wg}(\sigma)=\prod_{Q=0}^2 d^{n_Q}_Q \textrm{Wg}_{d_Q}(\sigma_Q)$ is a product of Weingarten functions in different charge sectors where $\sigma_Q$ is the permutation on the charge sector $Q$ implemented by $\sigma$ and where $d_{0}=d_{2}=d^2$ and $d_1=2d^2$. The legs in this tensor carry the charge indices $\boldsymbol{q}$ and the vertices carry the permutations $\sigma$. This naturally leads to an SM model on an an-isotropic hexagonal lattice shown in Fig. \ref{hex_lattice} with two types of edge in an analogous way to the non-symmetric case~\cite{Nahum_2018,Zhou_2019,Jian_2020,Li2021,Dias2022}.

The partition function for this model is given by a sum over the charge configurations on each edge and the permutations (compatible with the charges) at each vertex and where the vertical and diagonal edges have the following associated weights,
\begin{equation}
\includegraphics[width=0.07\textwidth, valign=c]{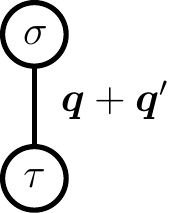} \to \textrm{Wg}(\sigma\tau^{-1})\sim \order{d^{-2|\sigma\tau^{-1}|}}, \quad \includegraphics[width=0.09\textwidth, valign=c]{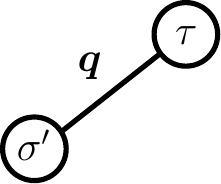} \to d^{-|\sigma'\tau^{-1}|},
\end{equation}
where $|\sigma|$ is the transposition distance of $\sigma$ (from $\mathbb{1}$). In the $d\to\infty$ limit, the Weingarten functions lock together the incoming and outgoing permutations $\sigma=\tau$. Together with the initial and final boundary conditions, $\sigma_0=\sigma_t=\mathbb{1}$, the permutation degrees of freedom are completely frozen, and the $n$-replica model decouples into $n$ independent discrete-time SEP chains for the charge degrees of freedom. Letting $d$ be large but finite allows different permutations to appear during the dynamics; domain-walls between different permutations $\sigma$ and $\tau$ have an energy cost of $\order{|\sigma\tau^{-1}|\log(d)}$ per unit length of domain wall.

\section{Effective Stochastic model at large $d$}

The dominant contributions to the SM partition function at large $d$ are the `all identity' configurations, in which the replicas are paired according to the identity permutation at every vertex. The next most relevant contributions will be given by dilute configurations of small domains of single transpositions in an identity background (see Fig. \ref{hex_lattice}). The smallest domains are the least costly, with a free energy contribution of $4\log(d)$, and are shown below,
\begin{equation}\label{bubble}
\includegraphics[width=0.09\textwidth, valign=c]{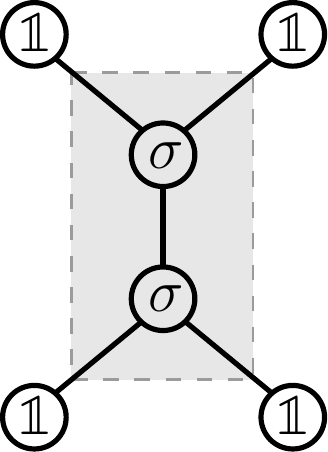} \quad \quad \includegraphics[width=0.09\textwidth, valign=c]{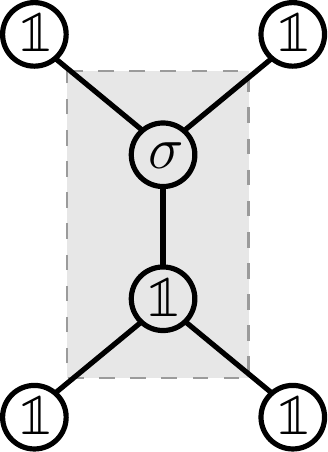} \quad \quad \includegraphics[width=0.09\textwidth, valign=c]{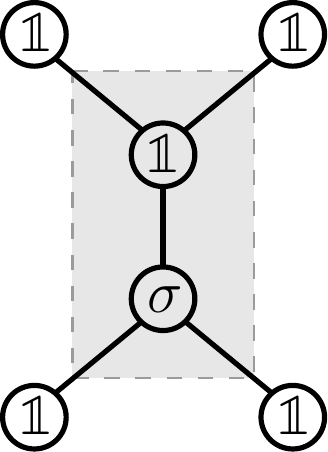},
\end{equation}
where $\sigma$ is a transposition of only two replicas and where the shaded region represents the two-site gate. These small domain contributions can also be counted in the brick-wall circuit picture by inserting projectors $P_{\mathbb{1}}$ onto the identity pairing space on each leg, or equivalently, by contracting the legs with states $\ket{\boldsymbol{q};\mathbb{1}}$. Upon doing this, we can replace $\overline{U^{\otimes n}\otimes U^{*\otimes n}}$ with a gate $G_{(n)}$ that explores only the $\sigma=\mathbb{1}$ subspace but has a modified charge dynamics,
\begin{equation}
\includegraphics[width=0.1\textwidth, valign=c]{projected_gate1}\ =\ \includegraphics[width=0.1\textwidth, valign=c]{projected_gate2} \ .
\end{equation}
Before giving an explicit expression for $G_{(n)}$, we adjust out notation for the local SM states with the identity permutation. Defining the states $\left|q\right) \equiv \frac{1}{\sqrt{d}}\sum_{a \in \mathcal{H}_q}\ket{aa^*} \in \mathcal{H}_{\textrm{loc}}\otimes\mathcal{H}^*_{\textrm{loc}}$, we can write $\ket{\boldsymbol{q};\mathbb{1}}= d^{n/2}\left|q_1\right)\otimes \cdots \otimes \left|q_n\right)$. Let us also define the two projectors $P$ and $K$ as follows,
\begin{equation}\label{PKdef}
P \equiv \frac{1}{2}\big[\left|0\right)\left|1\right) - \left|1\right)\left|0\right)\big]\big[\left(0\right|\left(1\right| - \left(1\right|\left(0\right|\big],\ K \equiv \mathbb{1} - P = \left|0\right)\left|0\right) \left(0\right|\left(0\right|+  \left|1\right)\left|1\right) \left(1\right|\left(1\right| + 
\frac{1}{2}\big[\left|0\right)\left|1\right) + \left|1\right)\left|0\right)\big]\big[\left(0\right|\left(1\right| + \left(1\right|\left(0\right|\big].
\end{equation}
Using superscripts to indicate in which replica the projectors act, the gate $G_{(n)}$ can be written as
\begin{equation}\label{modified gate}
G_{(n)} = \prod_{\alpha=1}^n K^{(\alpha)} + a(d) \sum_{ \alpha <\beta } P^{(\alpha)} P^{(\beta)} \prod_{\gamma \neq \alpha,\beta} K^{(\gamma)} + \order{d^{-8}},
\end{equation}
where $a(d)=[4d^4-1]^{-1}$.

\subsection{Sketch of derivation}
To see this result, we will multiply the averaged gate by SM states in the identity pairing space. We will consider a identity permutation state with charges $\boldsymbol{q}$ and $\boldsymbol{q}'$ on the incoming legs (one for each site) such that the total charges $\boldsymbol{Q}=\boldsymbol{q}+\boldsymbol{q}'$ satisfies $Q_m\leq Q_{m+1}$, i.e., a charge configuration in which the replicas are already collected into groups of like charge sector. The averaged gate then decomposes in an obvious way, $\overline{U^{\otimes n}\otimes U^{*\otimes n}}\ket{\boldsymbol{q};\mathbb{1}}\ket{\boldsymbol{q}';\mathbb{1}}=\prod_{Q=0}^2 \overline{U_Q^{\otimes n_Q}\otimes U_Q^{*\otimes n_Q}}\ket{\boldsymbol{q};\mathbb{1}}\ket{\boldsymbol{q}';\mathbb{1}}$. 

For the charge sectors $Q=0,2$, the charges on all legs are frozen (the charge on each site in each replica is the same). Dropping the charge labels from the states and retaining only the permutation label, the averaged $Q$-sector gate takes a simple form,
\begin{equation}
\overline{U_Q^{\otimes n_Q}\otimes U_Q^{*\otimes n_Q}} = \sum_{\sigma_Q,\tau_Q\in S_{n_Q}} \textrm{Wg}_{d_Q}(\sigma_Q\tau_Q^{-1})\ket{\sigma_Q}\bra{\tau_Q}.
\end{equation}
%Multiplying by an identity permutation state and noticing that the Weingarten function $d_Q^{n_Q}\textrm{Wg}_{d_Q}(\sigma_Q\tau_Q^{-1})$ is the matrix inverse of the Gram matrix $G_Q\equiv d^{-n_Q}\bra{\sigma_Q}\ket{\tau_Q} = d_Q^{-|\sigma_Q \tau_Q^{-1}|}$, we find
%\begin{equation}
%\overline{U_Q^{\otimes n_Q}\otimes U_Q^{*\otimes n_Q}}\ket{\mathbb{1}} = \sum_{\sigma_Q,\tau_Q\in S_{n_Q}}\textrm{Wg}_{d_Q}(\sigma_Q\tau_Q^{-1}) \ket{\tau_Q}\bra{\sigma_Q}\ket{\mathbb{1}} = \ket{\mathbb{1}}.
%\end{equation}
Now note that the identity permutation state is an eigenvector with eigenvalue $1$, since it corresponds to contracting each unitary with its conjugate in the same replica:
\begin{equation}
\overline{U_Q^{\otimes n_Q}\otimes U_Q^{*\otimes n_Q}}\ket{\mathbb{1}}  = \ket{\mathbb{1}}.
\end{equation}
Putting the charge labels back in, the restriction of $\overline{U_Q^{\otimes n_Q}\otimes U_Q^{*\otimes n_Q}}$ to the identity subspace is given by
\begin{equation}
\left(\overline{U_Q^{\otimes n_Q}\otimes U_Q^{*\otimes n_Q}}\right)_{\mathbb{1}} = d_Q^{-n_Q} \ket{(Q,\cdots,Q);\mathbb{1}}\bra{(Q,\cdots,Q);\mathbb{1}}.
\end{equation}
This can also be written graphically as a tensor diagram,
\begin{equation}
\left(\overline{U_Q^{\otimes n_Q}\otimes U_Q^{*\otimes n_Q}}\right)_{\mathbb{1}} = d_Q^{-n_Q} \includegraphics[width=0.2\textwidth, valign=c]{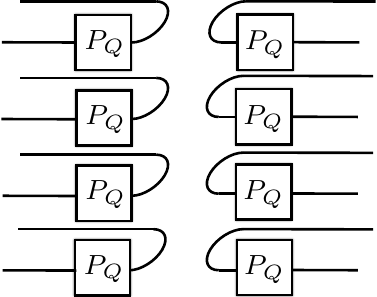},
\end{equation}
where $P_Q$ is the projector on charge sector $Q$. 

We now turn to the $Q=1$ sector, and make use of tensor diagrams again to write $\overline{U_1^{\otimes n_1}\otimes U_1^{*\otimes n_1}}$ as
\begin{equation}\label{sigma_tau_outer}
\overline{U_1^{\otimes n_1}\otimes U_1^{*\otimes n_1}} = \sum_{\sigma_1,\tau_1\in S_{n_1}}\textrm{Wg}_{d_1}(\sigma_1\tau_1^{-1}) \includegraphics[width=0.28\textwidth, valign=c]{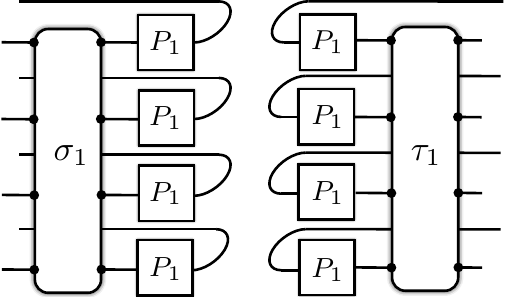},
\end{equation}
where $P_1$ is the projector onto the $Q=1$ subspace. Projecting this onto the identity permutation subspace is more complicated than in the previous cases. Unlike the $Q=0,2$ sectors where there is no freedom in the choice of charge configurations on the two sites, the $Q=1$ sector does have freedom. In particular, the allowed charge states on each of the two sites within each replica are $(q,q')=(0,1)$ and $(1,0)$. The identity permutation subspace (in the $Q=1$ sector) is spanned by the states $\ket{\boldsymbol{v}}$ defined diagrammatically below,
\begin{equation}
\ket{\boldsymbol{v}} \equiv d^{-n_1/2}_1 \includegraphics[width=0.1\textwidth, valign=c]{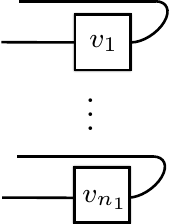},
\end{equation}
where $v_i \in \{P_1, Z\}$ are sums of projectors onto the charge configurations $(q,q') = (0,1)$ and $(1,0)$, $P_1=P_{01}+P_{10}$ and $Z=P_{01}-P_{10}$. The restriction of $\left(\overline{U_1^{\otimes n_1}\otimes U_1^{*\otimes n_1}}\right)_{\mathbb{1}}$ to the identity permutation space can be written in terms of these states,
\begin{equation}
\left(\overline{U_1^{\otimes n_1}\otimes U_1^{*\otimes n_1}}\right)_{\mathbb{1}} = \sum_{\boldsymbol{v},\boldsymbol{w}}C_{\boldsymbol{v},\boldsymbol{w}}\ket{\boldsymbol{v}}\bra{\boldsymbol{w}}.
\end{equation}
To determine the coefficient $C_{\boldsymbol{v},\boldsymbol{w}}$, we contract the tensor diagram in eq.~\eqref{sigma_tau_outer} by $\ket{\boldsymbol{w}}$ and $\bra{\boldsymbol{v}}$. It is not hard to check that the choice $\boldsymbol{w} = (P_1,\cdots,P_1)$ is in fact an eigenstate of $\overline{U_1^{\otimes n_1}\otimes U_1^{*\otimes n_1}}$ with eigenvalue $1$, and that any state with an odd number of $Z$'s will be annihilated due to the tracelessness of $Z$ and the fact that $Z$ is an involution (it is unavoidable that an odd number of $Z$'s will appear on the same loop in the tensor diagram). This leaves us with choice of $\boldsymbol{v}$ and $\boldsymbol{w}$ that have at least two $Z$'s. For such states, only the diagonal matrix elements $C_{\boldsymbol{v},\boldsymbol{v}}$ with $\boldsymbol{v}$ having only two insertions of $Z$ contribute at $\order{d^{-4}}$, all other matrix elements are $\order{d^{-8}}$. Concretely, for the state $\boldsymbol{v}$ with $Z$'s inserted in replicas $\alpha$ and $\beta$, we find $C_{\boldsymbol{v},\boldsymbol{v}}\approx d_Q^{-2}$, where the leading order contribution comes from $\sigma_1=\tau_1=T_{\alpha,\beta}$ (the transposition of replicas $\alpha$ and $\beta$). For $n=2$, this matrix element is calculated exactly to be $C_{\boldsymbol{v},\boldsymbol{v}} = a(d)$. To $\order{a(d)}$, we have the following for the restriction to the identity subspace,
\begin{equation}
\left(\overline{U_1^{\otimes n_1}\otimes U_1^{*\otimes n_1}}\right)_{\mathbb{1}} = d_1^{-n_1}  \includegraphics[width=0.2\textwidth, valign=c]{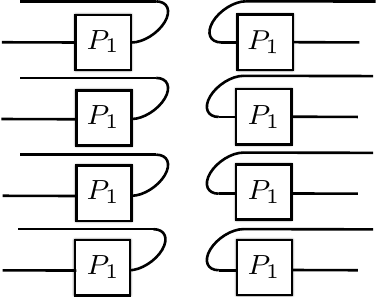} + a(d) d_1^{-n_1} \sum_{\alpha <  \beta} \ \ \includegraphics[width=0.2\textwidth, valign=c]{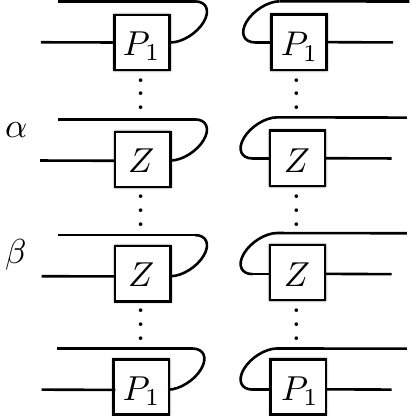} + \order{a(d)^2 d_1^{-n_1}}.
\end{equation}
The penultimate step is to notice that the projectors $P$ and $K$ defined in eq.~\eqref{PKdef} have the following diagrammatic form
\begin{equation}
P = d_1^{-1} \includegraphics[width=0.15\textwidth, valign=c]{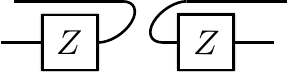},\quad
 K = d_0^{-1} \includegraphics[width=0.15\textwidth, valign=c]{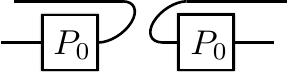} + d_1^{-1} \includegraphics[width=0.15\textwidth, valign=c]{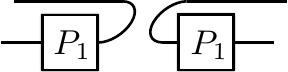} + d_2^{-1} \includegraphics[width=0.15\textwidth, valign=c]{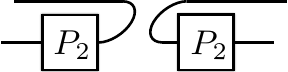}.
\end{equation}
The final step is to sum over all charge sector configurations (rather than the ordered configuration we first considered). Doing so recovers the gate $G_{(n)}$ as defined in eq.~\eqref{modified gate}.

\subsection{Transition probabilities for the $n=2$ replica stochastic model}

For the $n=2$ stochastic model, a pair of neighboring sites $x$ and $y$ are updated by $G_{(2)}$ with the transition probabilities given below. For the states in the charge sectors $(Q_1,Q_2)\neq (1,1)$, we have the transitions
\begin{align}
\ket{q_1, q_2}_x\ket{q_1,q_2}_y &\to \ket{q_1, q_2}_x\ket{q_1,q_2}_y, \nonumber \\
\ket{q_1, q_2}_x\ket{q_1,\overline{q}_2}_y &\to \frac{1}{2}\left(\ket{q_1, 0}_x\ket{q_1,1}_y + \ket{q_1, 1}_x\ket{q_1,0}_y\right),\nonumber \\
\ket{q_1, q_2}_x\ket{\overline{q}_1,q_2}_y &\to \frac{1}{2}\left(\ket{0,q_2}_x\ket{1,q_2}_y + \ket{1,q_2}_x\ket{0,q_2}_y\right),
\end{align}
for $q_1,q_2 \in \{0,1\}$ and where $\overline{q}\equiv 1- q$. For the charge sector $(Q_1,Q_2) = (1,1)$ we find that the transitions are modified by the parameter $a(d)$ in the following way,
\begin{align}
\ket{q, q}_x\ket{\overline{q},\overline{q}}_y &\to \frac{1+a}{4}\left(\ket{0,0}_x\ket{1,1}_y + \ket{1,1}_x\ket{0,0}_y\right) + \frac{1-a}{4}\left(\ket{0,1}_x\ket{1,0}_y + \ket{1,0}_x\ket{0,1}_y\right)\nonumber\\
\ket{q, \overline{q}}_x\ket{\overline{q},q}_y &\to \frac{1+a}{4}\left(\ket{0,1}_x\ket{1,0}_y + \ket{1,0}_x\ket{0,1}_y\right) + \frac{1-a}{4}\left(\ket{0,0}_x\ket{1,1}_y + \ket{1,1}_x\ket{0,0}_y\right).
\end{align}
%This effective model inherits an $n$-fold $SU(2)$ invariance (one for each chain) from the SM model, allowing for arbitrary rotations of the charge basis in each chain. For each choice of filling $f_i$ in the $i$-th chain, this effective stochastic process has a unique steady state $\prod_{i=1}^n \ket{\textrm{GS}_{f_i}}$, where $\ket{\textrm{GS}_{f_i}}$ is the equal weight superposition of states at filling $f_i$. The state $\ket{\textrm{GS}_{f}}$ is also the ground state of an isotropic Heisenberg ferromagnet in the appropriate spin sector (viewing the two-level system on each site as a spin-$1/2$ degree of freedom).

\section{Spin wave analysis}
The $n$-replica stochastic model evolves under the transfer matrix $T_{n}$, given by an `even-odd' and `odd-even' layer of two-site gates $G_{(n)}$ (eq.~\eqref{modified gate}). Expectation values of (time-ordered) two-time observables $A(t)B(0)$ in the $n$-replica model can be viewed as a vector overlap between observables in the replicated operator space,
\begin{equation}
\langle A(t)B(0) \rangle_{\textrm{$n$-replica}}\equiv \bra{A} T_n^{t} \ket{B \rho^{\otimes n}}.
\end{equation}
The circuit averaged variance $\overline{C}_2 \equiv \overline{\langle \Delta Q^2\rangle} - \overline{\langle \Delta Q\rangle^2}$ is given in our stochastic model by
\begin{equation}
\overline{C}_2 = \langle \Delta \hat{Q}_R^{(1) 2} \rangle_{\textrm{$1$-replica}} - \langle \Delta \hat{Q}_R^{(1)} \Delta \hat{Q}_R^{(2)} \rangle_{\textrm{$2$-replica}}.
\end{equation}
The first term agrees exactly with the SEP prediction as it is only a single replica quantity. Notice also, that after expanding $\Delta \hat{Q}_R = \hat{Q}_R(t)-\hat{Q}_R(0)$, all terms with one or fewer factor of $\hat{Q}_R(t)$ cancels with like terms in the corresponding expression for the SEP prediction $C^{\textrm{SEP}}_2$. The difference from SEP is then given by
\begin{align}
\Delta C_2 \equiv C^{\textrm{SEP}}_2 - \overline{C}_2 &= \langle \Delta \hat{Q}_R^{(1)} \Delta \hat{Q}_R^{(2)} \rangle_{\textrm{$2$-replica}} - \langle \Delta \hat{Q}_R^{(1)} \rangle_{\textrm{$1$-replica}}^2  \nonumber\\
&= \bra{\hat{Q}_R,\hat{Q}_R}\left(T_2(t) -T^{(1)}_1(t)T^{(2)}_1(t)\right)\ket{\rho,\rho}.
\end{align}
where $\ket{a,b} = \ket{a^{(1)}b^{(2)}}$. Define a rotated basis $\left| \uparrow\right) \equiv \frac{1}{\sqrt{2}} \left(  \left| 0 \right) + \left| 1\right) \right)$, $ \left| \downarrow\right) \equiv \frac{1}{\sqrt{2}} \left(  \left| 0 \right) - \left| 1\right) \right)$ (where the states $\left| 0 \right)$ and $ \left| 1 \right)$ were introduced in the previous appendix). Using these states we can write the local charge operator $Q_x$ as $\ket{Q_x} = \frac{\sqrt{2^L}}{2}\left(\ket{\textrm{GS}_0} - \ket{x}\right)$, where $\ket{\textrm{GS}_0}  = \left| \uparrow\cdots\uparrow\right)$ is the spin `up' polarized state and $\ket{x} =  \left| \uparrow\cdots\uparrow\downarrow_x\uparrow\cdots\uparrow\right)$ is the state with a single over-turned spin at site $x$. The charge operators for the left and right halves of the system are then given by $\ket{Q_R} = \frac{\sqrt{2^L}}{2}\left(\frac{L}{2}\ket{\textrm{GS}_0} - \ket{r}\right)$ and $\ket{Q_L} = \frac{\sqrt{2^L}}{2}\left(\frac{L}{2}\ket{\textrm{GS}_0} - \ket{l}\right)$, where $\ket{r}=\sum_{x\in \textrm{right}} \ket{x}$ and $\ket{l}=\sum_{x\in \textrm{left}} \ket{x}$. With these definitions, we write
\begin{equation}
\Delta C_2 = \frac{2^L}{4}\bra{r,r}\left(T_2(t) -T^{(1)}_1(t)T^{(2)}_1(t)\right)\ket{\rho,\rho}.
\end{equation}
The number of overturned spins in each replica is independently conserved by the transfer matrix. This enables us to select the component of the initial state that has a single overturned spin in each replica,

\begin{equation}\label{lr shorthand}
\Delta C_2 = \left(\frac{\tanh(\mu/2)}{2}\right)^2 \bra{r,r}\left(T_2(t) -T^{(1)}_1(t)T^{(2)}_1(t)\right)\ket{l-r,l-r}.
\end{equation}
The effective stochastic models have a global $SU(2)$ invariance in each of their chains, and have a single stationary state for a particular choice of fillings defined independently for each chain (the spins on each chain can have independently oriented $z$-axes when defining filling). This stationary state is given by the equal weight combinations of all spin configurations within a particular spin sector (filling) in each chain. If the state in a particular chain is a ground state of the ferromagnetic Heisenberg model, for instance $\ket{\textrm{GS}_1} = \ket{l} + \ket{r}$, this chain becomes inert, and no interactions between this chain and others take place. Therefore, $(T_2(t)-T^{(1)}_1(t)T^{(2)}_1(t))\ket{\textrm{GS}_1,\psi} = (T_1^{(2)}(t)-T_1^{(2)}(t))\ket{\textrm{GS}_1,\psi}=0$ for any $\ket{\psi}$. We use this fact to replace $\ket{l-r,l-r}$ by $4\ket{l,l}$, giving
\begin{equation}\label{C2 T short}
\Delta C_2 = \tanh(\frac{\mu}{2})^2 M(t), \quad M(t) \equiv \bra{r,r}\left(T_2(t) -T^{(1)}_1(t)T^{(2)}_1(t)\right)\ket{l,l}.
\end{equation}

Similar steps allow us to express any cumulant as a simple overlap of few-magnon states. In general, the $n$-th cumulant can be expressed as an overlap of states with $n$ magnons distributed among the $n$ chains of the $n$-chain stochastic model.

\subsection{Effective Hamiltonian}
By softening the gates $T_n$ of the discrete time stochastic models, we consider a continuous time stochastic process with an effective Hamiltonian $T^t_{n}\to T_n(t) \equiv e^{-t H_n}$. The gates are softened by taking $K\to e^{-\varepsilon P}$ and $P^{(\alpha)}P^{(\beta)}\to \varepsilon P^{(\alpha)}P^{(\beta)}$ and keeping only the $\order{\varepsilon}$ terms,
\begin{equation}
G_{(n),x,y} \to \mathbb{1} - \varepsilon\bigg(\sum_{\alpha} P_{x,y}^{(\alpha)} - a(d) \sum_{ \alpha < \beta } P_{x,y}^{(\alpha)}P_{x,y}^{(\beta)} \bigg) \approx e^{-\varepsilon \left(\sum_{\alpha} P_{x,y}^{(\alpha)} - a(d) \sum_{ \alpha < \beta } P_{x,y}^{(\alpha)}P_{x,y}^{(\beta)} \right)}.
\end{equation}
The effective Hamiltonian is then given by
\begin{equation}
H_n = \sum_j \sum_{\alpha} P_{j,j+1}^{(\alpha)} - a(d) \sum_j \sum_{ \alpha < \beta } P_{j,j+1}^{(\alpha)}P_{j,j+1}^{(\beta)}.
\end{equation}
The stationary states of the discrete time processes are the ground states of the effective Hamiltonians $H_n$. By studying the low-energy properties of $H_2$, we will determine the long-time behaviour of the charge transfer variance $\overline{C_2}$. Writing \eqref{C2 T short} for $\Delta C_2$ using the effective Hamiltonian evolution, we have
\begin{equation}\label{C2 short}
\Delta C^H_2 = \tanh(\frac{\mu}{2})^2 M(t), \quad M(t) \equiv \bra{r,r}\left(e^{-t H_2} -e^{-t(H^{(1)}_1+H^{(1)}_2)}\right)\ket{l,l}.
\end{equation}
Restricted to the subspace of a single overturned spin in each chain (which we refer to as the $1+1$ space), the Hamiltonian $H_2$ becomes,
\begin{equation}
H_{2,1+1} = \frac{1}{2}\sum_{x,y}\left[\left(\ket{x,y}-\ket{x+1,y}\right)\left(\bra{x,y}-\bra{x+1,y}\right) + \left(\ket{x,y}-\ket{x,y+1}\right)\left(\bra{x,y}-\bra{x,y+1}\right)\right] -\sum_x a \ket{v_{x}}\bra{v_{x}},
\end{equation}
where $\ket{v} = \frac{1}{2}\left(\ket{x,x}-\ket{x+1,x}-\ket{x,x+1}+\ket{x+1,x+1}\right)$. The inter-chain interaction is only activated when the overturned spin in each chain are distance $|x-y|\leq 1$ apart. This means that the low lying interactions are magnons (spin-waves) that propagate independently in each chain except when they experience a contact interaction. In the center of mass frame, this interaction takes the form of a scattering potential at the origin. The eigenstates of the Hamiltonian are therefore scattering states, with the same energies as in the model without a scatter impurity (in the thermodynamic limit). It is easy to check that the anti-symmetric scatter state with momenta $(k_1, k_2)$, $\ket{k_1,k_2}-\ket{k_2,k_1}$, is an eigenstate of $H_{2,1+1}$ with eigenvalue $E_2(k_1,k_2) = 2- \cos(k_1)-\cos(k_2) = E_1(k_1) + E_1(k_2)$. The anti-symmetric scattering states are also eigenstates of the Hamiltonian without an inter-chain interaction term, and so the contributions from these states in eq.~\eqref{C2 short} cancel. Labelling the symmetric scattering eigenstates as $\ket{k_1,k_2}'_+$ and $\ket{k_1,k_2}_+$ for the Hamiltonian with and without the scattering impurity respectively, we write
\begin{equation}
M(t) = \frac{1}{2}\sum_{x_1,x_2\in L}\sum_{y_1,y_2\in L} \sum_{k_1,k_2} e^{-t E(k_1,k_2)}\left(\psi'_+(k_1,k_2;y_1,y_2)\psi'_+(k_1,k_2;x_1,x_2)^*+\psi_+(k_1,k_2;y_1,y_2)\psi_+(k_1,k_2;x_1,x_2)^*\right),
\end{equation}
where the factor of $1/2$ accounts for the double counting momenta and where $\psi'_+(k_1,k_2;x_1,x_2) = \bra{x_1,x_2}\ket{k_1,k_2}'_+$ and $\psi_+(k_1,k_2;x_1,x_2) = \bra{x_1,x_2}\ket{k_1,k_2}_+$ take the following form
\begin{equation}\label{ansatz}
\psi'_+(k_1,k_2;x_1,x_2) = \frac{1}{L}e^{ip(x_1+x_2)}\phi'_+(p,k;x_2-x_1),\quad \phi'_+(p,k;r) = \frac{1}{\sqrt{2}} \left(e^{ik|r|}+e^{i\theta(p,k) - ik|r|} + \alpha(p,k)\delta(r)\right),
\end{equation}
with $p=k_1+k_2$ and $k=k_2-k_1$. The scattering phase shift $\theta(p,k)$ and the function $\alpha(p,k)$ at the origin vanish in the absence of a scattering potential. To determine $\theta$ and $\alpha$, we use the fact that $H_{2,1+1}$ is diagonal in $p$. Defining the real-space difference coordinate $r = x_2-x_1$, the total momentum $p$ sector Hamiltonian is given by
\begin{equation}
H_p = \sum_r \left(2\ket{r}\bra{r} - h\ket{r}\bra{r+1} - h\ket{r+1}\bra{r}\right) - \frac{a}{4}\left(2h\ket{0}-\ket{1}-\ket{-1}\right)\left(2h\bra{0}-\bra{1}-\bra{-1}\right),
\end{equation}
where $h(p) \equiv \cos(p)$. Using the ansatz $\ket{k}'_+ = \frac{1}{L}\sum_r \phi'_+(p,k;r) \ket{r}$, with  $\phi'_+(p,k;r)$ given by eq.~\eqref{ansatz}, we find that the scatter phase shift and $\alpha$ are given at small momenta by
\begin{equation}
e^{i\theta}\approx\frac{(2+a)k-ia(k^2-p^2)^2/4}{(2+a)k+ia(k^2-p^2)^2/4},\quad \Re{\alpha} \approx -\frac{a}{2}\left(p^2-k^2\right), \quad \Im{\alpha}\sim ik.
\end{equation}
Converting the sums to integrals and then using the fact that $e^{-t E(k_1,k_2)}$ suppresses all large momentum contributions to expand in small momenta, $E(k_1,k_2)=2(1-\cos(p)\cos(k))\approx p^2+k^2$ as well as to extend the momentum integrals to $(-\infty,\infty)$, we find
\begin{equation}
M(t) = \frac{1}{16\pi^2}\int_L d^2x \int_R d^2y \int dk \ dp\  e^{-t(p^2+k^2)}e^{ipw}\left(\phi'_+(p,k;\Delta_y)\phi'_+(p,k;\Delta_x)^*+\phi_+(p,k;\Delta_y)\phi_+(p,k;\Delta_x)^*\right),
\end{equation}
where $\Delta_x = x_2-x_1$ and likewise for $\Delta_y$ and where $w=y_1+y_2-x_1-x_2$. Expanding $\phi'_+$ and $\phi_+$ yields $M(t)=M_1(t)+M_2(t)+M_3(t)$ where each contributions is given by
\begin{equation}
M_i(t) = \frac{1}{32\pi^2}\int_L d^2x \int_R d^2y \int dk \ dp\  e^{-t(p^2+k^2)}e^{ipw} g_1(p,k,\Delta_x,\Delta_y),
\end{equation}
and where the $g_i$ are given by
\begin{align}
g_1(p,k,\Delta_x,\Delta_y) &= e^{i k (|\Delta_x| + |\Delta_y|)}(e^{-i\theta}-1) + c.c,\\
g_2(p,k,\Delta_x,\Delta_y) &= \alpha^*\left(e^{i k |\Delta_x|}+e^{i k |\Delta_y|}\right)\delta(\Delta_x) + \alpha\left(e^{-i k |\Delta_x|}+e^{-i k |\Delta_y|}\right)\delta(\Delta_y),\\
g_3(p,k,\Delta_x,\Delta_y) &= |\alpha|^2\delta(\Delta_x)\delta(\Delta_y).
\end{align}
We next rescale the momenta by $p\to p'=p\sqrt{t}$ and $k\to k'=k\sqrt{t}$ and the positions by $x_i\to x'_i=x_i /\sqrt{t}$ and $y_i\to x'_i=y_i /\sqrt{t}$. The third contribution $M_3(t)$ is sub-leading, at $\order{1/t}$, whereas the contributions $M_1(t)$ and $M_2(t)$ are given to leading order in $1/t$ by
\begin{align}
M_1(t) &= -\frac{1}{32\pi^2\sqrt{t}}\int_L d^2 x \int_R d^2 y \int dk \int dp\ e^{-p^2-k^2 - ipw}\frac{a(k^2-p^2)^2}{(2+a)k}\sin(k(|\Delta_x|+|\Delta_y|)),\\
M_2(t) &= \frac{1}{32\pi^2\sqrt{t}}\int_L d^2 x \int_R d^2 y \int dk \int dp\ e^{-p^2-k^2 - ipw}4\Re{\alpha}\cos(k|\Delta_x|)\delta(\Delta_y).
\end{align}
Discarding $\order{a^2}$ contributions, each of these integrals is equal to $a(d)/\left[16\sqrt{\pi t}\right]$. Putting this together gives a cumulant difference $\Delta C^H_2$ as
\begin{equation}\label{theory}
\Delta C^H_2(\mu,t) \approx \frac{a(d)\tanh(\mu/2)^2}{16\sqrt{\pi t}}.
\end{equation}
We test this result numerically and find excellent agreement with the theoretical prediction.
\begin{figure}[H]
\centering
\includegraphics[width=0.45\textwidth]{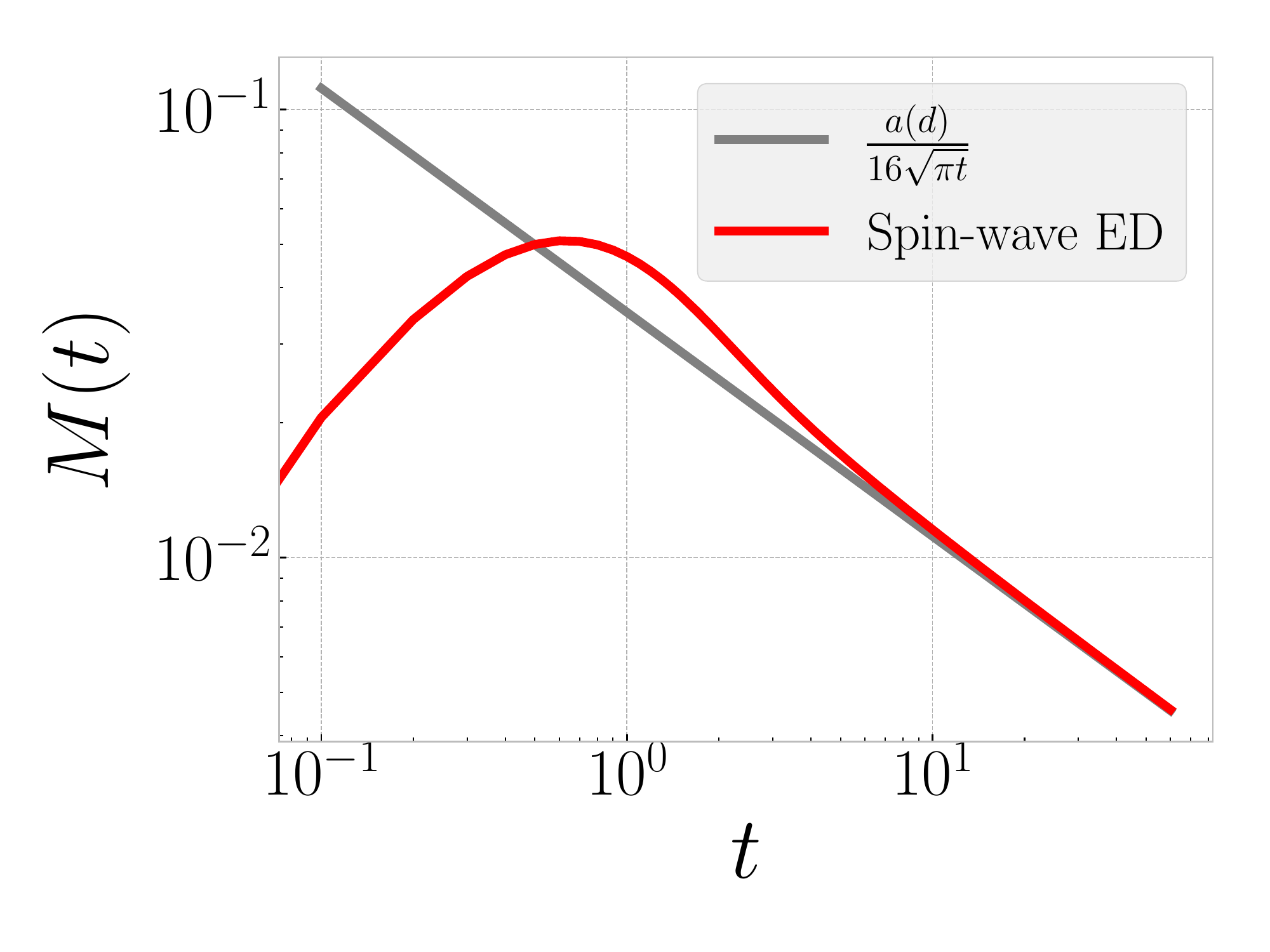}
\caption{$M(t) = \Delta C_2(t)/(a (d) \tanh(\mu)^2)$ is calculated numerically using spin-wave exact diagonalization data (red) and its late-time behavior is given by eq.~\eqref{theory} (grey).}
\end{figure}

In the linear-response regime ($\mu\ll 1$), we can leverage this result to find a theoretical prediction for the difference between SEP and the effective stochastic models in the third cumulant $\Delta C^H_3(\mu,t)$,
\begin{equation}
    \Delta C^H_3(\mu,t) \approx \frac{3\mu M(t)}{4} \approx \frac{3 a(d) \mu}{64\sqrt{\pi t}}.
\end{equation}

\section{Ab-initio matrix product state numerics}

\begin{figure}[H]
\centering
\includegraphics[width=0.45\textwidth]{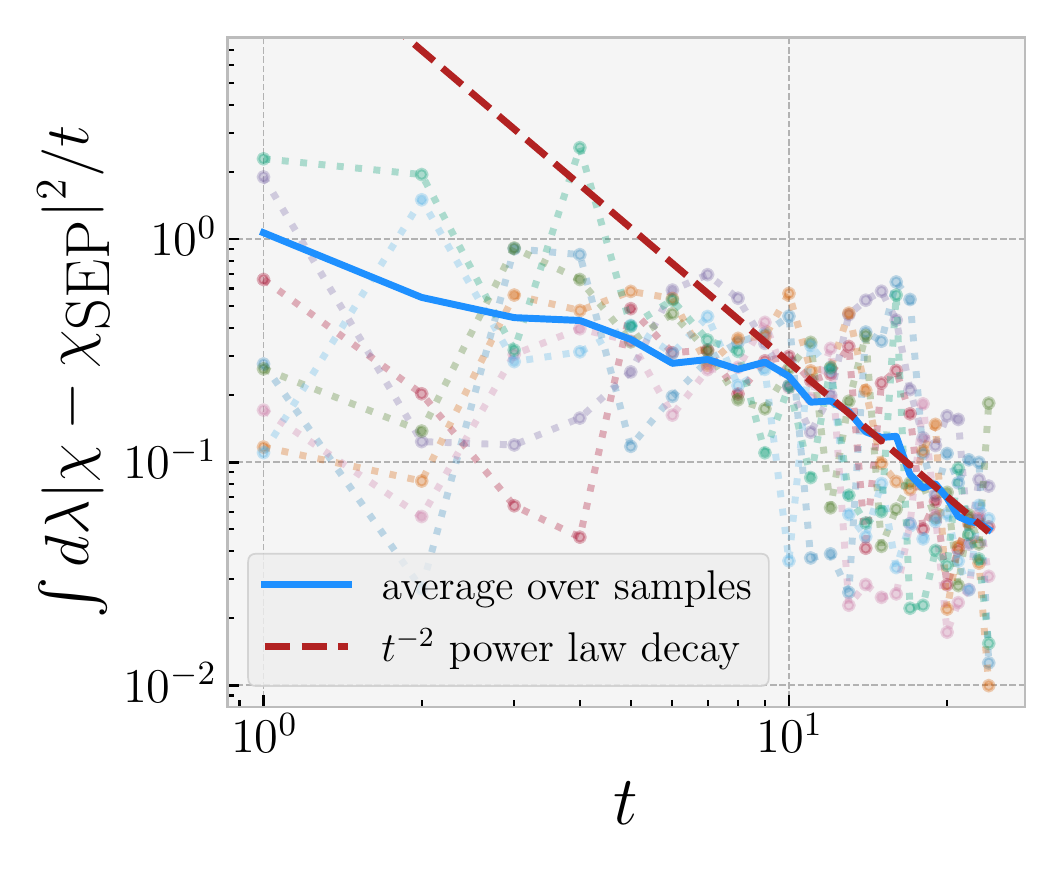}
\caption{Circuit-to-circuit fluctuations of the CGF for individual circuits (multicolored) and averaged over 35 circuit realisations (blue) using TEBD numerics.}
\label{circuit-to-circuit deviations}
\end{figure}

We verify the predictions of our effective statistical model by directly computing the FCS of individual realizations of random $U(1)$ quantum circuits using matrix-product state techniques. We focus on the case $q=2$ (qubits) for simplicity. Recall that the unitary evolution is generated by a brick-wall pattern of Haar-random 2-qubit charge-conserving gates (written in the charge basis $\ket{11},\ket{10},\ket{01},\ket{00}$)
\begin{equation}
U_{j,j+1}^{(t)} = \begin{pmatrix}
        1 & 0 & 0& 0 \\ 
        0 & e^{i(\alpha_{j,t}+\psi_{j,t})}\sqrt{1-\xi_{j,t}} & e^{i(\alpha_{j,t}+\chi_{j,t})} \sqrt{\xi_{j,t}} & 0 \\ 
        0 & -e^{i(\alpha_{j,t}-\chi_{j,t})}\sqrt{\xi_{j,t}} & e^{i(\alpha_{j,t}-\psi_{j,t})} \sqrt{1-\xi_{j,t}} & 0 \\
        0 & 0 & 0 & e^{{i\rho_{j,t}}}
    \end{pmatrix},
\end{equation}
where the  variables $\rho_{j,t}, \psi_{j,t}, \chi_{j,t}, \alpha_{j,t} \sim U(0, 2\pi)$ and $\xi_{j,t} \sim U(0, 1)$ are random in both space and time. In order to compute FCS, we introduce a ``counting field''~\cite{Touchette_2009,Esposito_2009,Tang_2014,Ridley2018,Kilgour2019,PhysRevLett.123.200601,PhysRevLett.128.090605,Erpenbeck2021,Popovic2021}, and modify the gates acting on the central bond in the system by adding phase factors ${e}^{\pm i \lambda/2}$ on the off-diagonal elements keeping track of charge transfer across that bond: 
\begin{equation}
U^{(t)}_{0,1}(  \lambda) = \begin{pmatrix}
        1 & 0 & 0& 0 \\ 
        0 & e^{i(\alpha_{0,t}+\psi_{0,t})}\sqrt{1-\xi_{0,t}} & e^{i(\alpha_{0,t}+\chi_{0,t})} \sqrt{\xi_{0,t}}  {e}^{ i \lambda/2} & 0 \\ 
        0 & -e^{i(\alpha_{0,t}-\chi_{0,t})}\sqrt{\xi_{0,t}}  {e}^{- i \lambda/2} & e^{i(\alpha_{0,t}-\psi_{0,t})} \sqrt{1-\xi_{0,t}} & 0 \\
        0 & 0 & 0 & e^{{i\rho_{0,t}}}
    \end{pmatrix}.
\end{equation}
Denoting the global unitary implementing this modified circuit up to time $t$ by $U(t,\lambda)$, the moment generating function of charge transfer can be computed from the overlap
\begin{equation} \label{eqCountingfield}
\langle e^{i \lambda Q }\rangle = {\Tr} \left(U^\dagger(t,-\lambda) U(t,\lambda) \rho_0 \right),
\end{equation}
where $\rho_0$ is the initial density matrix. (In equilibrium and at half-filling, we have $\rho_0 = \mathbb{1}/2^L$.) We can evaluate eq.~\eqref{eqCountingfield} using standard matrix-product state techniques and using the TEBD algorithm to compute the time evolution. We do this with bond dimension $\chi=2000$ to compute the CGF for individual circuit realisations, as shown in Fig. \ref{measurement cartoon}, finding that the circuit-to-circuit fluctuations around the SEP CGF diminish over time. These deviations are captured by the quantity $|\chi(\lambda,t)-\chi_{\textrm{SEP}}(\lambda,t)|^2/t$, where $\chi_{\textrm{SEP}}$ is the asymptotic SEP CGF (eq. \eqref{Derrida}). We integrate over the counting field to obtain a single measure of circuit-to-circuit fluctuations. This is shown in Fig. \ref{circuit-to-circuit deviations} below for eight individual circuits and the average over $n=35$ circuit realisations, $\int d\lambda \overline{|\chi(\lambda,t)-\chi_{\textrm{SEP}}(\lambda,t)|}^2/t$. We find a power law decay of $t^{-2}$, corresponding with a cumulant generating function $\chi(\lambda,t)/\sqrt{t} \sim \chi_{\textrm{SEP}}(\lambda,t)/\sqrt{t} + \order{1/t}$.

\section{Additional numerical data for the statistical mechanics model}
\begin{figure}[H]
\centering
  \raisebox{-0.5\height}{\includegraphics[width=0.4\textwidth]{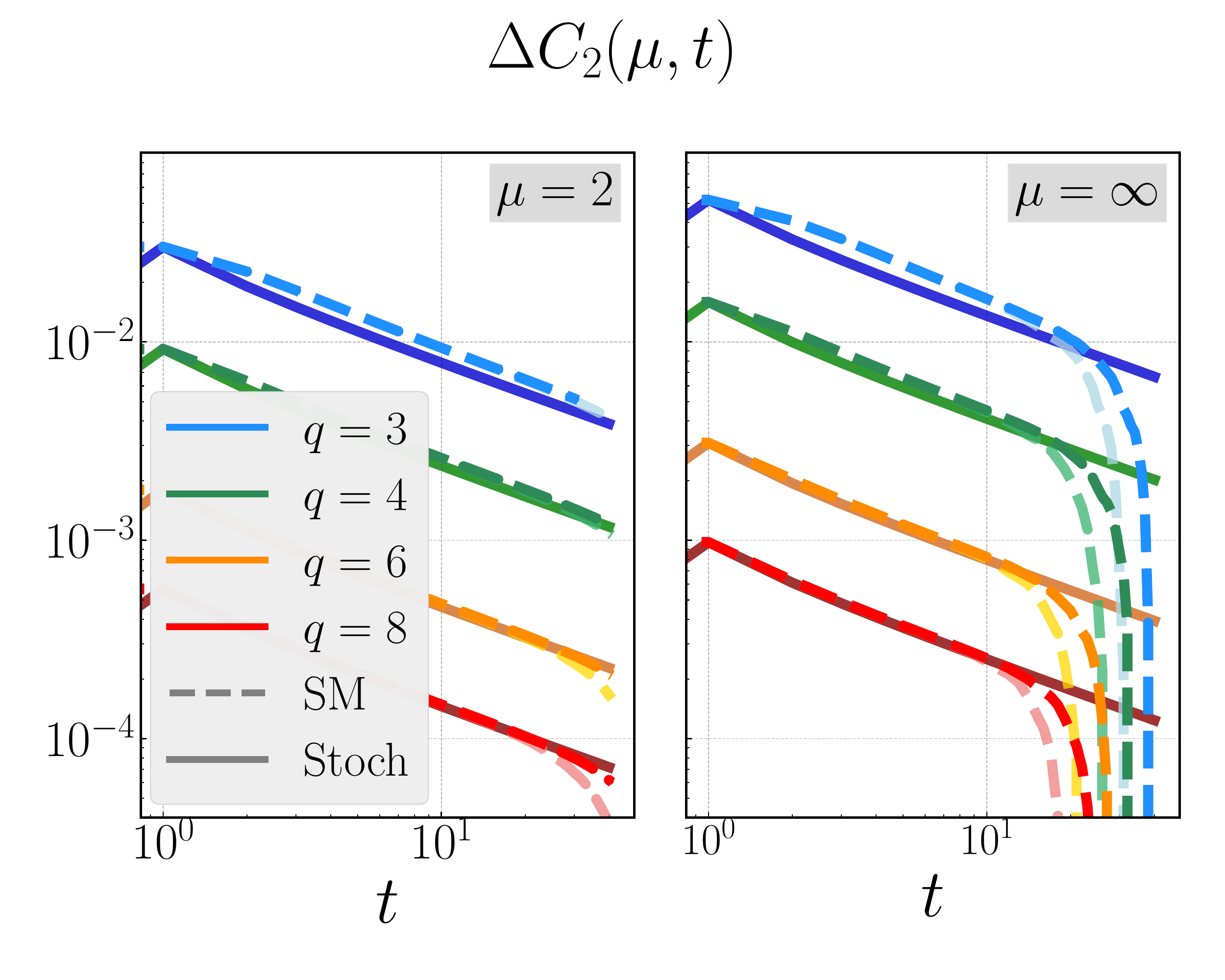}}
  \hspace{-2mm}\raisebox{-0.5\height}{\includegraphics[width=0.4\textwidth]{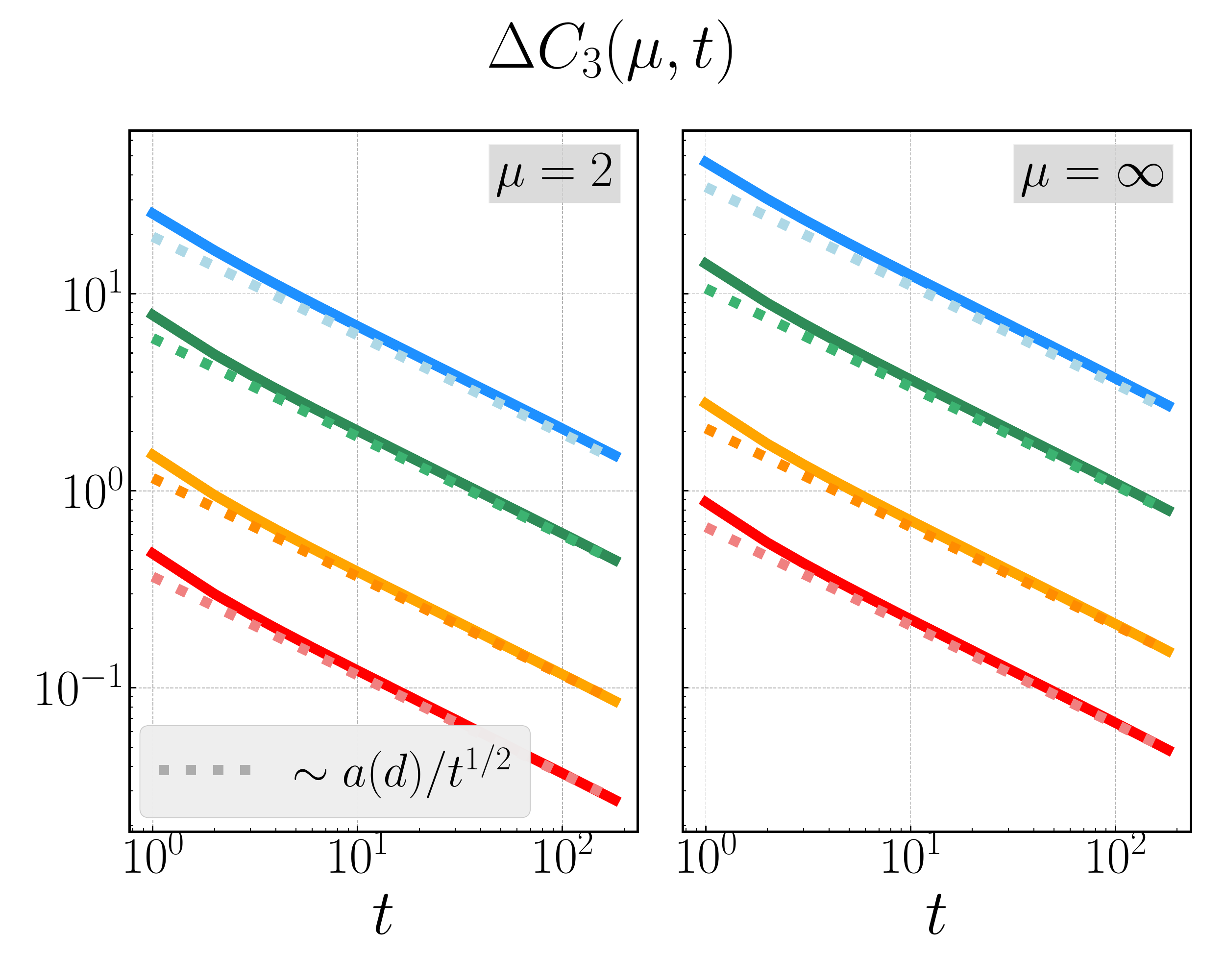}}
  \caption{The approach of circuit averaged cumulants $\overline{C}_n$(t) to their SEP values: (a) $\Delta C_2(t)$ data for $\mu=2,\infty$ using the statistical mechanics model (dashed) and the effective stochastic model (bold); (b) $\Delta C_3(t)$ data for $\mu=2,\infty$ using the effective stochastic model.}
\label{additional cumulants}
\end{figure}

In this appendix, we present cumulant data for initial biases not shown in the main text, $\mu=2,\infty$. Using TEBD with bond dimensions $\chi=1000$ and $\chi=1500$ we apply the $n=2$ SM transfer matrix, finding a $t^{-1/2}$ approach to the SEP value (while the data remains converged). We also verify this with the effective stochastic model which agrees remarkably well with the SM data. We also compute the third cumulant using the stochastic model, again finding a $t^{-1/2}$ deviation from SEP. The SM and stochastic model data is shown in Fig. \ref{additional cumulants}. 

%\bibliographystyle{apsrev4-1} % Tell bibtex which bibliography style to use
%\bibliography{referencesv2}

\end{document}